\documentclass[usenatbib]{mnras}
\usepackage{graphicx}   
\usepackage{amsmath}    

\title[Binary fraction determination method by the integrated spectrum]{Determination method of binary fractions by the integrated spectrum}
\author[F. Zhang, L. Li, Z. Han, et al.]{
F.~Zhang,$^{1,2,4}$\thanks{E-mail: zhangfh@ynao.ac.cn}
L.~Li,$^{1,2,4}$
Z.~Han$^{1,2,4}$
and X.~Wang$^{1,2,3}$
\\ 
$^{1}$Yunnan Observatories, Chinese Academy of Sciences, 396 Yangfangwang, Guandu District, Kunming, 650216, P. R. China\\
\tiny $^{2}$Key Laboratory for the Structure and Evolution of Celestial Objects, Chinese Academy of Sciences, 396 Yangfangwang, Guandu District, Kunming, 650216, P. R. China\\
$^{3}$University of Chinese Academy of Sciences, Beijing, 100049, China\\
$^{4}$ International Centre of Supernovae, Yunnan Key Laboratory, Kunming 650216, P. R. China 
}

\date{Accepted XXX. Received YYY; in original form ZZZ}
\pubyear{2023}

\begin{document}
\label{firstpage}
\pagerange{\pageref{firstpage}--\pageref{lastpage}}
\maketitle

\begin{abstract}
We need to resolve the individual stars for binary fraction determinations of stellar systems. Therefore, it is not possible to obtain the binary fractions for dense or distant stellar systems.
We proposed a method to determine the binary fraction of a dense or distant stellar system. The method is to first determine the binary fraction variation for any two adjacent regions and then add up those binary fraction variations along the radial direction to obtain the binary fraction for a stellar system. 
Binary fraction variation is derived by using ten binary fraction-sensitive spectral absorption feature indices (SAFIs) and the binary fraction variation calibrations in terms of these SAFIs.
Using this method, we first presented the binary fraction variations for twenty-one Galactic globular clusters (GCs). By comparisons, we find that they agree well with the binary fractions based on the main-sequence fiducial line method by previous studies. This verifies that the above mentioned method is feasible. 
Next, we presented the binary fraction variations of thirteen Galactic GCs.
We gave the relationships between binary fraction and various parameters, and found that binary fraction is negatively correlated with $N_{\rm HB}$ and $N_{\rm RR}$, binary fraction of some studies is not strongly correlated with $N_{\rm BS}$, and the number of GCs with large binary fraction is greater at extreme blue horizontal branch population ratio.
At last, if we want to obtain more accurate binary fraction, we suggest that the spectroscopic and photometric observations are conducted at an appropriate area interval for a stellar system.
\end{abstract}


\section{Introduction}
\label{sec:intro} 
Binary star systems are very fascinating members. Where did they form? How were they formed? These issues have always attracted the attention of many astronomers \citep{2013ARA&A..51..269D}.
Determining the amount of binary stars (or binary fraction) present in different stellar systems plays an important role in understanding the above issues. 
Double compact objects, a type of binary star, are important observation sources of LIGO, Virgo, KAGRA and LISA gravitational wave detectors (\citealt{2019PhRvX...9c1040A, 2008ApJS..174..223B, 2023LRR....26....2A}).
In the evolutionary population synthesis models with binary stars, binary fraction would affect the results of the stellar populations \citep{Zha04, 2016MNRAS.462.3302E}.
Therefore, the determination of binary fraction has important influences on binary star formation, predictions about gravitational wave events, and evolutionary population synthesis models with binary stars.

\citet{1935PASP...47...15K} first argued that empirically determining the multiplicity frequency and distributions of key orbital parameters among main-sequence stars would be highly valuable from a theoretical standpoint. Up to now, there are three main methods for binary fraction determinations of stellar systems (see the description of \citealt[][hereafter M12]{2012A&A...540A..16M}).
The first one is to perform spectroscopic observation(s) on a candidate binary star system, judge whether it is a binary star system through Doppler shift in the spectrum, and accumulate the number of binary stars to determine the binary fraction in a stellar system.
The second one is to conduct multiple photometric observations on a candidate binary star system, determine whether it is a photometric binary star system through light curve, and accumulate the number of binary stars to determine binary fraction.
The last one is to determine the number of stars located on the red side of the main-sequence fiducial line in an image (called as MSFL method) to get binary fraction.

All of the above binary fraction determination methods require very high spatial resolution. In addition, the first method also needs very high spectral resolution.
The modern large-scale surveys (such as TESS, ZTF, Gaia, and LAMOST, \citealt{2022ApJS..258...16P, 2020ApJ...905L...7B, 2024AN....34530123J, 2014ApJ...788L..37G}) have greatly improved the efficiency of determining binary fractions, but overall, the first and second methods are still relatively time-consuming.
Therefore, the binary fractions of only $\sim$70 Galactic globular clusters (GCs) have been studied. 
However, \citet[][Paper I]{2023A&A...679A..27Z} presented the calibrations of variation in binary fraction with a mass ratio of $q$>0.5 in terms to ten binary fraction-sensitive integrated spectral absorption feature index (SAFI) variations. 
In this way, we can use two or more integrated spectral energy distributions to determine the binary fractions for those stellar systems in which the individual stars cannot be resolved (such as distant or dense star clusters).

The outline of the paper is as follows. 
In Section 2, we describe the proposed method for the binary fraction determination, samples of GCs, data, and calculations.
In Section 3, we test the feasibility of some steps in the proposed method by comparing the derived binary fractions with those based on the MSFL method by M12 and \citet[][hereafter J15]{2015ApJ...807...32J} for twenty-one Galactic GCs, analyze the reasons for the differences, and give the suggestions for observations in order to determine binary fractions and obtain more accurate calibrations of binary fraction variation.
In Section 4, we present the binary fraction changes for thirteen  Galactic GCs, and discuss the relationships between binary fraction and various parameters and processes in conjunction with the results of M12 and J15. 
Finally, we present a summary in Section 5.

\section{Method, sample, data and calculations}
\label{sec:metsamdatcal} 
\begin{table*}
\caption{Calibration coefficient {$a_k$(=$\Delta$SAFI$_k/\Delta$$f_k$}, from Eq.\,9 of Paper I) between binary fraction difference and binary fraction-sensitive SAFI difference at the resolution of 5\,\AA.}
\leftline{
\begin{tabular}{l ccccc ccccc}
\hline
1 & 2 & 3 & 4 & 5 & 6 & 7 & 8  & 9  & 10 & 11 \\
   & Ca4455  & C$_2$4668  & TiO$_1$   & OIII-1  & OIII-2   & H$_{\beta}$   & H$_{\rm \delta_A}$ & H$_{\rm \gamma_A}$ & H$_{\rm \delta_F}$   & H$_{\rm  \gamma_F}$ \\
\hline
$a_k$   &   $-1.30$  & $-76.87$  & $-2.08$ & $-40.92$ &  $-35.14$ & $10.88$ & $125.64$ & $108.58$ & $101.35$ & $94.57$ \\
\hline
\end{tabular}
}
\label{Tab:slope}
\end{table*}

\begin{table*}
\caption{Main characteristics of the twenty-one GCs in Sample A. }
{
\begin{tabular}{lll rrr rrr rrr rr}
\hline
1 & 2 & 3 & 4 & 5 & 6 & 7 & 8  & 9  & 10 & 11 & 12 & 13 & 14 \\
NAME \quad (PCC AC) & $f$  & spectrum   & $N_{\rm HB}$ & $N_{\rm RR}$ & $N_{\rm BS}$ & $c$ & $R_{\rm C}$ & $R_{\rm HM}$ & $M_{\rm V}$ & [Fe/H] & HBR & $R_{\rm sun}$ & $R_{\rm GC}$ \\
   & reference   & reference   &    &    &    &    & ($\arcmin$) & ($\arcmin$) &      &     &   & (kpc) & (kpc) \\
\hline
NGC104 \ \ (...\,\,\,\,\,\, ...\,\,\,) & M12 J15 & ...\,\,\,\,\,\, S05 (1) U17 & $ 363 $ & $ 2 $ & $ 54 $   &  $      2.07  $   &  $      0.36  $   &  $      3.17  $   &  $     -9.42  $   &  $     -0.72  $   &  $     -0.99  $   &  $      4.50  $   &  $      7.40  $ \\
NGC1851 (...\,\,\,\,\,\,  AC) & M12 J15 & ...\,\,\,\,\,\, S05 (1) U17 & $ 307 $ & $ 28 $ & $ 81 $   &  $      1.86  $   &  $      0.09  $   &  $      0.51  $   &  $     -8.33  $   &  $     -1.18  $   &  $     -0.36  $   &  $     12.10  $   &  $     16.60  $ \\
NGC2298 (...\,\,\,\,\,\, AC) & M12 ...\,\,\,\,\,\, & ...\,\,\,\,\,\, S05 (2) ...\,\,\,\,\,\, & $ ... $ & $ ... $ & $ ... $   &  $      1.38  $   &  $      0.31  $   &  $      0.98  $   &  $     -6.31  $   &  $     -1.92  $   &  $      0.93  $   &  $     10.80  $   &  $     15.80  $ \\
NGC3201 (...\,\,\,\,\,\, ...\,\,\,) & M12 ...\,\,\,\,\,\, & ...\,\,\,\,\,\, S05 (2) U17 & $ 24 $ & $ 8 $ & $ 18 $   &  $      1.29  $   &  $      1.30  $   &  $      3.10  $   &  $     -7.45  $   &  $     -1.59  $   &  $      0.08  $   &  $      4.90  $   &  $      8.80  $ \\
NGC5904 (...\,\,\,\,\,\, AC) & M12 J15 & ...\,\,\,\,\,\, S05 (2) U17 & $ 172 $ & $ 43 $ & $ 28 $   &  $      1.73  $   &  $      0.44  $   &  $      1.77  $   &  $     -8.81  $   &  $     -1.29  $   &  $      0.31  $   &  $      7.50  $   &  $      6.20  $ \\
NGC5927 (...\,\,\,\,\,\, ...\,\,\,) & M12 J15 & P02 S05 (4) U17 & $ 219 $ & $ 1 $ & $ 23 $   &  $      1.60  $   &  $      0.42  $   &  $      1.10  $   &  $     -7.81  $   &  $     -0.49  $   &  $     -1.00  $   &  $      7.70  $   &  $      4.60  $ \\
NGC5986 (...\,\,\,\,\,\, ...\,\,\,) & M12 ...\,\,\,\,\,\, & ...\,\,\,\,\,\, S05 (1) U17 & $ 269 $ & $ 15 $ & $ 49 $   &  $      1.23  $   &  $      0.47  $   &  $      0.98  $   &  $     -8.44  $   &  $     -1.59  $   &  $      0.97  $   &  $     10.40  $   &  $      4.80  $ \\
NGC6121 (...\,\,\,\,\,\, ...\,\,\,) & M12 J15 & ...\,\,\,\,\,\, S05 (1) U17 & $ ... $ & $ 2 $ & $ ... $   &  $      1.65  $   &  $      1.16  $   &  $      4.33  $   &  $     -7.19  $   &  $     -1.16  $   &  $     -0.06  $   &  $      2.20  $   &  $      5.90  $ \\
NGC6171 (...\,\,\,\,\,\, ...\,\,\,) & M12 ...\,\,\,\,\,\, & ...\,\,\,\,\,\, S05 (2) U17 & $ 39 $ & $ 9 $ & $ 33 $   &  $      1.53  $   &  $      0.56  $   &  $      1.73  $   &  $     -7.12  $   &  $     -1.02  $   &  $     -0.73  $   &  $      6.40  $   &  $      3.30  $ \\
NGC6218 (...\,\,\,\,\,\, ...\,\,\,) & M12 J15 & P02 S05 (1) U17 & $ 41 $ & $ ... $ & $ 31 $   &  $      1.34  $   &  $      0.79  $   &  $      1.77  $   &  $     -7.31  $   &  $     -1.37  $   &  $      0.97  $   &  $      4.80  $   &  $      4.50  $ \\
NGC6254 (...\,\,\,\,\,\, ...\,\,\,) & M12 ...\,\,\,\,\,\, & ...\,\,\,\,\,\, S05 (1) U17 & $ ... $ & $ 1 $ & $ ... $   &  $      1.38  $   &  $      0.77  $   &  $      1.95  $   &  $     -7.48  $   &  $     -1.56  $   &  $      0.98  $   &  $      4.40  $   &  $      4.60  $ \\
NGC6352 (...\,\,\,\,\,\, ...\,\,\,) & M12 J15 & ...\,\,\,\,\,\, S05 (1) U17 & $ ... $ & $ ... $ & $ ... $   &  $      1.10  $   &  $      0.83  $   &  $      2.05  $   &  $     -6.47  $   &  $     -0.64  $   &  $     -1.00  $   &  $      5.60  $   &  $      3.30  $ \\
NGC6362 (...\,\,\,\,\,\, ...\,\,\,) & M12 J15 & ...\,\,\,\,\,\, S05 (1) U17 & $ 42 $ & $ 6 $ & $ 25 $   &  $      1.09  $   &  $      1.13  $   &  $      2.05  $   &  $     -6.95  $   &  $     -0.99  $   &  $     -0.58  $   &  $      7.60  $   &  $      5.10  $ \\
NGC6624 (PCC ...\,\,\,) & M12 J15 & P02 S05 (2) U17 & $ 123 $ & $ 1 $ & $ 45 $   &  $      2.50  $   &  $      0.06  $   &  $      0.82  $   &  $     -7.49  $   &  $     -0.44  $   &  $     -1.00  $   &  $      7.90  $   &  $      1.20  $ \\
NGC6637 (...\,\,\,\,\,\, ...\,\,\,) & M12 J15 & P02 S05 (1) U17 & $ 156 $ & $ 3 $ & $ 48 $   &  $      1.38  $   &  $      0.33  $   &  $      0.84  $   &  $     -7.64  $   &  $     -0.64  $   &  $     -1.00  $   &  $      8.80  $   &  $      1.70  $ \\
NGC6652 (...\,\,\,\,\,\, ...\,\,\,) & M12 J15 & ...\,\,\,\,\,\, S05 (2) U17 & $ 61 $ & $ 1 $ & $ 45 $   &  $      1.80  $   &  $      0.10  $   &  $      0.48  $   &  $     -6.66  $   &  $     -0.81  $   &  $     -1.00  $   &  $     10.00  $   &  $      2.70  $ \\
NGC6723 (PCC ...\,\,\,) & M12 J15 & ...\,\,\,\,\,\, S05 (1) U17 & $ 106 $ & $ 24 $ & $ 24 $   &  $      1.11  $   &  $      0.83  $   &  $      1.53  $   &  $     -7.83  $   &  $     -1.10  $   &  $     -0.08  $   &  $      8.70  $   &  $      2.60  $ \\
NGC6752 (PCC ...\,\,\,) & M12 J15 & ...\,\,\,\,\,\, S05 (1) U17 & $ ... $ & $ 7 $ & $ ... $   &  $      2.50  $   &  $      0.17  $   &  $      1.91  $   &  $     -7.73  $   &  $     -1.54  $   &  $      1.00  $   &  $      4.00  $   &  $      5.20  $ \\
NGC7078 (PCC ...\,\,\,) & M12 J15 & ...\,\,\,\,\,\, S05 (2) U17 & $ 399 $ & $ 81 $ & $ 46 $   &  $      2.29  $   &  $      0.14  $   &  $      1.00  $   &  $     -9.19  $   &  $     -2.37  $   &  $      0.67  $   &  $     10.40  $   &  $     10.40  $ \\
NGC7089 (...\,\,\,\,\,\,  AC) & M12 ...\,\,\,\,\,\, & ...\,\,\,\,\,\, S05 (1) U17 & $ 184 $ & $ 12 $ & $ 15 $   &  $      1.59  $   &  $      0.32  $   &  $      1.06  $   &  $     -9.03  $   &  $     -1.65  $   &  $      0.96  $   &  $     11.50  $   &  $     10.40  $ \\
NGC2808 (...\,\,\,\,\,\,  AC) & ...\,\,\,\,\,\, J15 & ...\,\,\,\,\,\, S05 (2) U17 & $ 854 $ & $ 15 $ & $ 107 $   &  $      1.56  $   &  $      0.25  $   &  $      0.80  $   &  $     -9.39  $   &  $     -1.14  $   &  $     -0.49  $   &  $      9.60  $   &  $     11.10  $ \\
\hline
\end{tabular}
}
\label{Tab:smpA-mc}
\end{table*}

\begin{table*}
\caption{Main characteristics of the thirteen GCs in Sample B. Same as Table~\ref{Tab:smpA-mc} but without the references for the binary fraction data.}
\begin{tabular}{llr rrr rrr rrr r}
\hline
1 & 2 & 3 & 4 & 5 & 6 & 7 & 8 & 9 & 10 & 11 & 12 & 13  \\
NAME\ \ \ \ (PCC AC) & spectrum  & $N_{\rm HB}$ & $N_{\rm RR}$ & $N_{\rm BS}$ & $c$ & $R_{\rm C}$ & $R_{\rm HM}$ & $M_{\rm V}$ & [Fe/H] & HBR & $R_{\rm sun}$ & $R_{\rm GC}$ \\
   &  reference  &    &    &    &    & (\arcmin) & (\arcmin) &      &     &   & (kpc) & (kpc) \\
\hline
NGC1904 (PCC  AC) & ...\,\,\,\,\,\, S05 (2) U17 & $ 172 $ & $ 4 $ & $ 39 $   &  $      1.70  $   &  $      0.16  $   &  $      0.65  $   &  $     -7.86  $   &  $     -1.60  $   &  $      0.89  $   &  $     12.90  $   &  $     18.80  $ \\
NGC6266 (PCC ...\,\,\,\,\,\,) & ...\,\,\,\,\,\, S05 (1) U17 & $ 504 $ & $ 74 $ & $ 49 $   &  $      1.71  $   &  $      0.22  $   &  $      0.92  $   &  $     -9.18  $   &  $     -1.18  $   &  $      0.32  $   &  $      6.80  $   &  $      1.70  $ \\
NGC6284 (PCC ...\,\,\,\,\,\,) & P02 S05 (2) U17 & $ 135 $ & $ 2 $ & $ 22 $   &  $      2.50  $   &  $      0.07  $   &  $      0.66  $   &  $     -7.96  $   &  $     -1.26  $   &  $    -99.00  $   &  $     15.30  $   &  $      7.50  $ \\
NGC6304 (...\,\,\,\,\,\, ...\,\,\,\,\,\,) & ...\,\,\,\,\,\, S05 (1) U17 & $ 105 $ & $ ... $ & $ 53 $   &  $      1.80  $   &  $      0.21  $   &  $      1.42  $   &  $     -7.30  $   &  $     -0.45  $   &  $     -1.00  $   &  $      5.90  $   &  $      2.30  $ \\
NGC6316 (...\,\,\,\,\,\, ...\,\,\,\,\,\,) & ...\,\,\,\,\,\, S05 (2) U17 & $ 214 $ & $ ... $ & $ ... $   &  $      1.65  $   &  $      0.17  $   &  $      0.65  $   &  $     -8.34  $   &  $     -0.45  $   &  $     -1.00  $   &  $     10.40  $   &  $      2.60  $ \\
NGC6333 (...\,\,\,\,\,\, ...\,\,\,\,\,\,) & ...\,\,\,\,\,\, S05 (1) U17 & $ ... $ & $ ... $ & $ ... $   &  $      1.25  $   &  $      0.45  $   &  $      0.96  $   &  $     -7.95  $   &  $     -1.77  $   &  $      0.87  $   &  $      7.90  $   &  $      1.70  $ \\
NGC6342 (PCC ...\,\,\,\,\,\,) & ...\,\,\,\,\,\, S05 (2) U17 & $ 66 $ & $ ... $ & $ 30 $   &  $      2.50  $   &  $      0.05  $   &  $      0.73  $   &  $     -6.42  $   &  $     -0.55  $   &  $     -1.00  $   &  $      8.50  $   &  $      1.70  $ \\
NGC6356 (...\,\,\,\,\,\, ...\,\,\,\,\,\,) & P02 S05 (1) U17 & $ 389 $ & $ ... $ & $ 71 $   &  $      1.59  $   &  $      0.24  $   &  $      0.81  $   &  $     -8.51  $   &  $     -0.40  $   &  $     -1.00  $   &  $     15.10  $   &  $      7.50  $ \\
NGC6522 (PCC ...\,\,\,\,\,\,) & ...\,\,\,\,\,\, S05 (1) U17 & $ 138 $ & $ ... $ & $ ... $   &  $      2.50  $   &  $      0.05  $   &  $      1.00  $   &  $     -7.65  $   &  $     -1.34  $   &  $      0.71  $   &  $      7.70  $   &  $      0.60  $ \\
NGC6528 (...\,\,\,\,\,\, ...\,\,\,\,\,\,) & P02 S05 (4) U17 & $ ... $ & $ ... $ & $ ... $   &  $      1.50  $   &  $      0.13  $   &  $      0.38  $   &  $     -6.57  $   &  $     -0.11  $   &  $     -1.00  $   &  $      7.90  $   &  $      0.60  $ \\
NGC6553 (...\,\,\,\,\,\, ...\,\,\,\,\,\,) & P02 S05 (1) U17 & $ ... $ & $ ... $ & $ ... $   &  $      1.16  $   &  $      0.53  $   &  $      1.03  $   &  $     -7.77  $   &  $     -0.18  $   &  $     -1.00  $   &  $      6.00  $   &  $      2.20  $ \\
NGC6569 (...\,\,\,\,\,\, ...\,\,\,\,\,\,) & P02 S05 (1) U17 & $ 192 $ & $ ... $ & $ 52 $   &  $      1.31  $   &  $      0.35  $   &  $      0.80  $   &  $     -8.28  $   &  $     -0.76  $   &  $    -99.00  $   &  $     10.90  $   &  $      3.10  $ \\
NGC6626 (...\,\,\,\,\,\, ...\,\,\,\,\,\,) & P02 S05 (1) ...\,\,\,\,\,\, & $ ... $ & $ 5 $ & $ ... $   &  $      1.67  $   &  $      0.24  $   &  $      1.97  $   &  $     -8.16  $   &  $     -1.32  $   &  $      0.90  $   &  $      5.50  $   &  $      2.70  $ \\
\hline
\end{tabular}
\label{Tab:smpB-mc}
\end{table*}

\begin{table*}
\caption{Weight $w_k$ of binary fraction-sensitive SAFI for the calculation of the binary fraction difference (see Eq.~\ref{Eq:method-1step1}).}
\leftline{
\begin{tabular}{ l lllll lllll}
\hline
1 & 2 & 3 & 4 & 5 & 6 & 7 & 8  & 9  & 10 & 11  \\
  & Ca4455  & C$_2$4668  & TiO$_1$   & OIII-1  & OIII-2   & H$_{\beta}$   & H$_{\rm \delta_A}$ & H$_{\rm \gamma_A}$ & H$_{\rm \delta_F}$   & H$_{\rm  \gamma_F}$ \\
\hline
$w_k$ & 0.040 & 0.096  & 0.155  & 0.203  & 0.166  & 0.041 & 0.07 & 0.04  & 0.1052  & 0.079\\
\hline
\end{tabular}
}
\label{Tab:weight}
\end{table*}

\begin{figure}
\centering
\includegraphics[height=5.7cm, width=6.00cm]{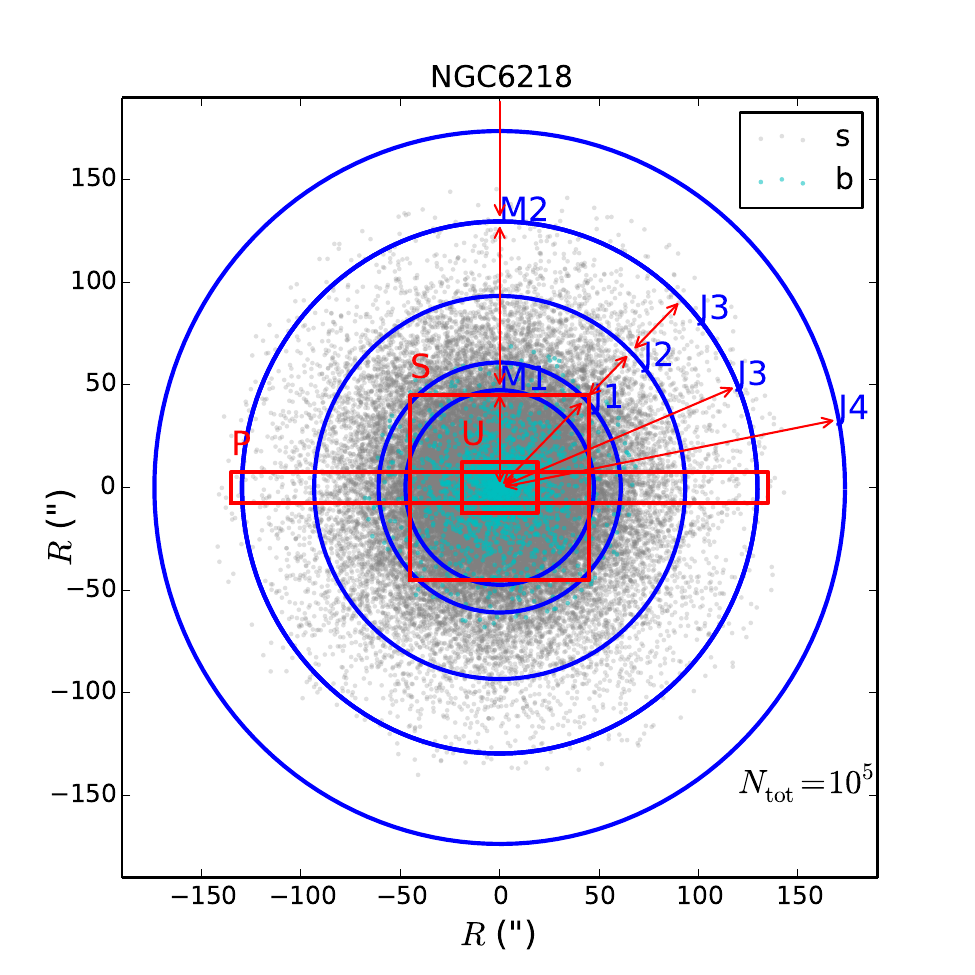}
\caption{The spatial distribution of simulated single (gray points) and binary stars (cyan points) in NGC6218 (comprising $N_{\rm tot}$=$10^5$ stars), as well as the binary fraction observed (blue circles and red arrows, M1, M2, J1, ..., J5 are for the 1st and 2nd of M12, 1st to 5th regions of J15) and spectrum observed regions (red rectangles, P, S and U are for the regions of P02, S05 and U17).
M3 region is not fully included due to the small scope of this figure.
}
\label{Fig:paperI}
\end{figure}

\begin{figure}
\centering
\includegraphics[height=5.7cm, width=6.00cm]{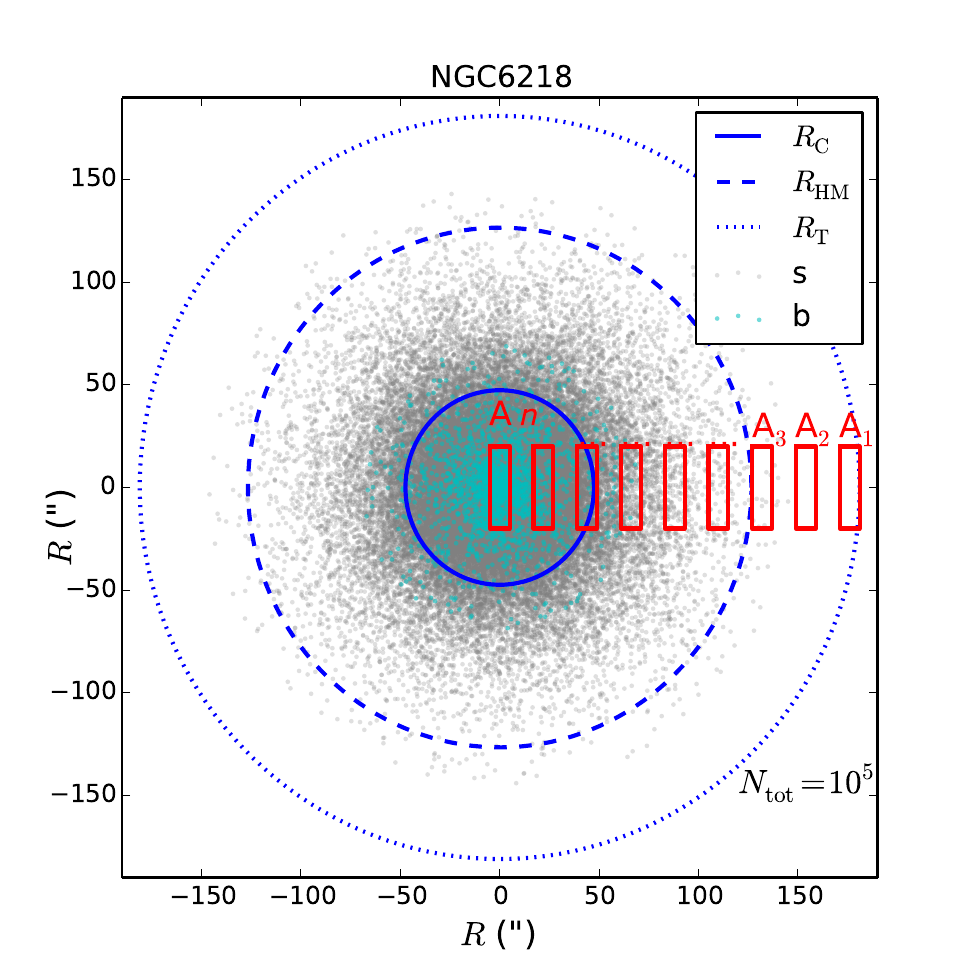}
\caption{The spectrum observed regions (red rectangles) for the method proposed in this work, also shown are the core radius $R_{\rm C}$, half mass radius $R_{\rm HM}$, and tidal radius $R_{\rm T}$ (blue solid, dashed and dotted lines). The gray and cyan points have the same meanings as Fig.~\ref{Fig:paperI}.}
\label{Fig:method}
\end{figure}

\begin{figure}
\leftline{
\includegraphics[height=6.0cm, width=8.0cm]{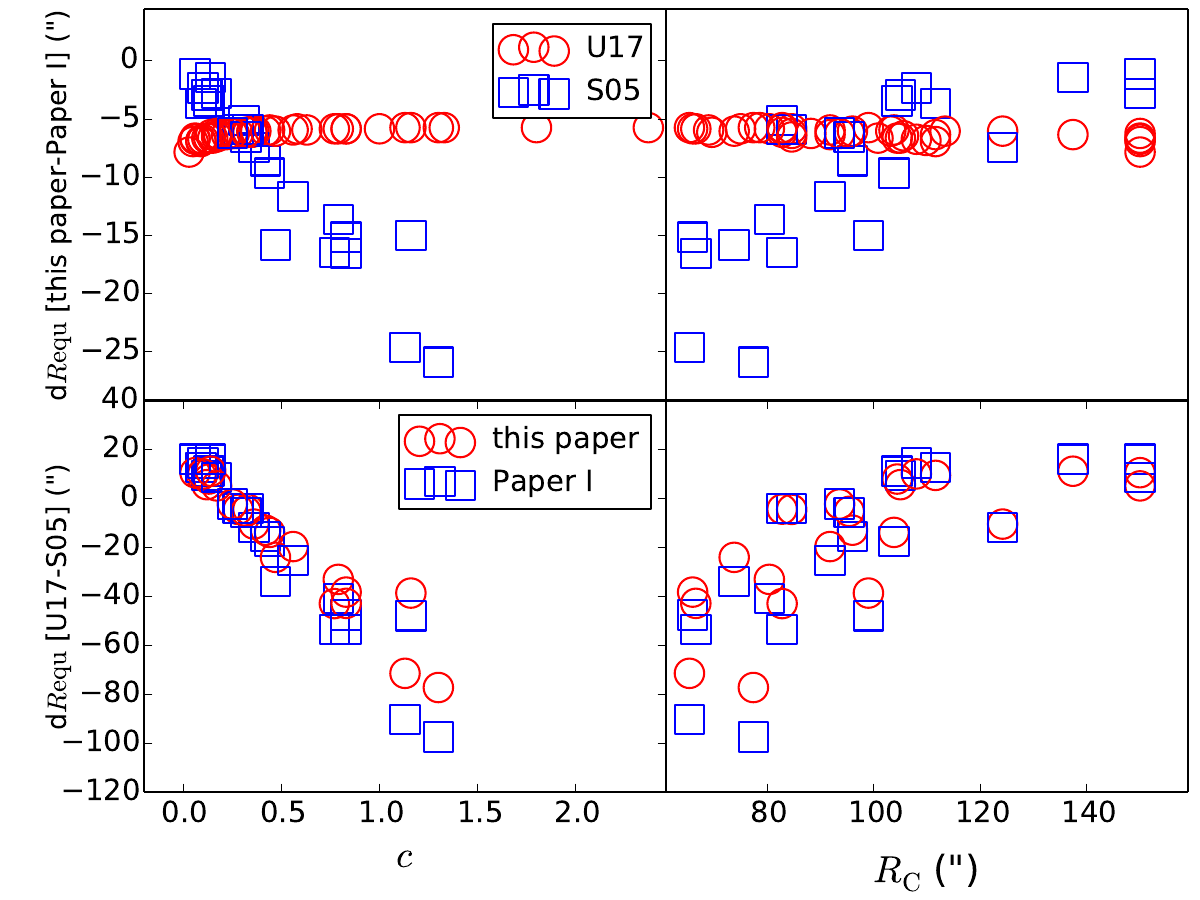}
}
\caption{{\bf Top panels:} Difference in the equivalent radius d$R_{\rm equ}$[this paper$-$Paper I] as functions of the central concentration $c$ (left) and core radius $R_{\rm C}$ (right) for the forty-four GCs in the U17 (red circles) and S05 spectrum samples (blue squares). {\bf Bottom panels:} Difference in the equivalent radius d$R_{\rm equ}$[U17$-$S05] as functions of $c$ (left) and $R_{\rm C}$ (right) for this work (red circles) and Paper I (blue squares).}
\label{Fig:fig2}
\end{figure}

\subsection{Method for the binary fraction determination}
\label{subsec:method} 
In Paper I, we established the calibration of binary fraction variation in terms of SAFI variation.
In order to briefly describe the method for the determining calibration relationships, in Fig.~\ref{Fig:paperI} we first present the spatial distribution of simulated single and binary stars (the total $N$=$10^5$) in NGC6218, as well as the binary fraction observed regions and spectrum observed regions.
The simulated single and binary stars are obtained based on the radial binary fraction profile given in Paper I and the surface density profile of the \citet{1962AJ.....67..471K} model,
\begin{equation}
\begin{split}
P_{\rm sd}   &=   \Bigl\{{\frac{1}{[1+(R/R_{\rm C})^2]^{\frac{1}{2}}}}-{\frac{1}{[1+(R_T/R_{\rm C})^2]^{\frac{1}{2}}}}\Bigl\}^2 , \\
&= \Bigl\{{\frac{1}{[1+(R/R_{\rm C})^2]^{\frac{1}{2}}}}-{\frac{1}{[1+(e^c)^2]^{\frac{1}{2}}}}\Bigl\}^2 , 
\end{split}
\label{Eq:psd}
\end{equation}
where $c$, $R_{\rm T}$ and $R_{\rm C}$ are the central concentration, tidal and core radii, and the values are from the \citet{1996AJ....112.1487H} catalog.
The binary fraction observed regions are circular or annular, in this figure we use red arrows to represent the inner and outer radii of the ring (blue circles), where M1 and M2 are the two observed regions of M12, and J1,..., J5 are the first to fifth areas of \citet[][hereafter J15]{2015ApJ...807...32J}. 
The spectrum observed areas are rectangular (red), with the regions of \citet[][hereafter P02]{2002A&A...395...45P}, \citet[hereafter U17]{2017MNRAS.468.3828U} and \citet[][hereafter S05]{2005ApJS..160..163S} represented by P, S and U to the upper left corner of the rectangles. 
The method for establishing the calibrations of binary fraction variation in terms of SAFI variation is as follows.

We constructed the radial binary fraction profile by using the binary fraction data within circular or annular regions (the blue circles and red arrows in Fig.~\ref{Fig:paperI}) and the surface density profile of the \citet{1962AJ.....67..471K} model.
Transforming the spectrum observed region (the red rectangles of Fig.~\ref{Fig:paperI}) to the circular one with an equivalent radius $R_{\rm equ}$, we obtained the binary fraction $f_{i,j}$ within the $j$-th spectrum observed region and the binary fraction difference $\Delta$$f_{i}$ (=$f_{i,j}$-$f_{i,j'}$, $j$$\ne$$j'$) between/among different regions for the $i$-th GC.
In the transformation of equivalent radius $R_{\rm equ}$ in Paper I, its minimal value is equal to smaller one between the width and height of the rectangle region. If we use larger one between width and height as the equivalent radius, it is obviously unreasonable for very narrow areas (such as the P02 region, see Fig.~\ref{Fig:paperI}). Therefore, we used half of the diagonal of a rectangle observed region as the equivalent radius for the S05 and U17 observations. This method includes two parameters of width and height and is better than the above two methods. 
For the P02 observations, we used 2.5 times slit width as the equivalent radius because any an integrated spectrum was constructed based on three long-slit spectra (one within the nuclear region and two within the adjacent fields twice the slit width from the center).

Subsequently, in Paper I we analyzed the relationship between the difference in the $k$-th spectral index $\Delta$SAFI$_{i,k}$ (=SAFI$_{i,k,j}$$-$SAFI$_{i,k,j'}$, $j$$\ne$$j'$) between/among different regions for the $i$-th GC and the $k$-th SAFI range covered by all GCs, and found that ten SAFIs (Ca4455, C$_2$4668, TiO$_1$, OIII-1, OIII-2, H$_\beta$, H$_{\rm \delta A}$,  H$_{\rm \gamma A}$, H$_{\rm \delta F}$ and H$_{\rm \gamma F}$) are binary fraction sensitive.
At last, we presented the calibrations of binary fraction variation in terms of these ten SAFI variations by fitting the values of all GCs (Eq.\,9 of Paper I). The calibrations are as follows.
\begin{equation}
\Delta f_k= \frac{1}{a_k} \Delta {\rm SAFI}_k  \ \ \  (k\le10),
\label{Eq:calib}
\end{equation}
where $a_k$ is the calibration coefficient for the $k$-th SAFI and is given in Table.~\ref{Tab:slope}.
In the calibrations, $\Delta f$ is independent of the SAFI type, but in the calculations the derived binary fraction is related to the SAFI type. Therefore, in the above expression $\Delta f$ uses the subscript $k$.
The binary fractions by using a mass ratio of $q$>0.5 are relatively accurate compared to the other binary fraction types, and the integrated spectral energy distributions or SAFIs at the resolutions of 5\,\AA\ and Lick/IDS system are more sensitive to binary fraction than at low resolution. Therefore, the calibrations presented in Paper I are based on differences in binary fraction with $q$>0.5 and SAFI at the resolution of  5\,\AA.
It is worth mentioning that in the following text, unless otherwise specified, binary fraction refers to the binary fraction with $q$>0.5.

In this paper, we propose the following method to obtain the binary fraction in a stellar system (illustrated by Fig.\,~\ref{Fig:method}).
The first step is to obtain the first intermediate- or high-resolution integrated spectral energy distribution within a region, located at about a tidal radius $R_{\rm T}$ from the center (A$_1$ region in Fig.~\ref{Fig:method}) and with a binary fraction $f_{\rm A1}$$\simeq$0.
The second step is to obtain the second integrated spectral energy distribution by appropriately moving the observed region inward (A$_2$ region in Fig.~\ref{Fig:method}) and derive its binary fraction $f_{\rm A2}$ through the SAFIs at the resolution of 5\,\AA\,(Section~\ref{subsec:transformation}), the calibration coefficients (\{$a_k$\}) in Table~\ref{Tab:slope} (see Eq.~\ref{Eq:calib}), and weights ($\{w_k\}$) for the ten SAFIs (Section~\ref{subsec:weights}),
\begin{equation}
f_{\rm A2}  = f_{\rm A1}+\Delta f_{\rm A1},  
\label{Eq:method-1step3}
\end{equation}
and
\begin{equation}
\Delta f_{\rm A1}  =  \Sigma_{k=1}^{10}  \Delta f_{{\rm A1},k}\,w_k, \\ 
\label{Eq:method-1step1}
\end{equation}
where $\Delta f_{\rm A1}$ is the binary fraction difference, and $\Delta f_{{\rm A1},k}$ is the binary fraction difference derived from the $k$-th index difference between A1 and A2 regions.
Substituting into Eq.~\ref{Eq:calib},  the above expressions become,
\begin{equation}
\Delta f_{\rm A1}  =  \Sigma_{k=1}^{10}  \frac{\Delta {\rm SAFI}_{{\rm A1},k}}{a_k} \,w_k, 
\label{Eq:method-1step4}
\end{equation}
and
\begin{equation}
f_{\rm A2} = f_{\rm A1} + \Sigma_{k=1}^{10}  \frac{\Delta {\rm SAFI}_{{\rm A1},k}}{a_k} \,w_k, \\
\label{Eq:method-1step2}
\end{equation}
where $\Delta {\rm SAFI}_{{\rm A1},k}$ is the $k-$th SAFI difference between A1 and A2 regions.
Similar to step 2, the third to $n$-th steps are to obtain the third to $n$-th integrated spectral energy distributions (within A$_3$, ..., A$n$ regions in Fig.\ref{Fig:method}) and derive the third to $n$-th binary fractions  ($f_{\rm A3}$, ..., $f_{\rm An}$) by continuously moving the observed region inwards until its center covers the cluster's center (see Fig.~\ref{Fig:method}).
The last step is to add up these values to obtain the binary fraction for a stellar system,
\begin{equation}
f=\frac{\Sigma \ f_{\rm An}\, N_{{\rm A}n}} {\Sigma \ N_{{\rm A}n}}, 
\label{Eq:method}
\end{equation}
where $N_{{\rm A}n}$ is the (relative) number in the A$n$ region and can be obtained from Eq.~\ref{Eq:psd}. If a GC has the distribution of spectrum regions (red rectangles) as shown in Fig.~\ref{Fig:paperI}, we can only obtain its approximate binary fraction to verify one step of the proposed method (Section 3), or we can obtain its approximate binary fraction if its binary fractions has not been studied before (Section 4).

\subsection{Samples of GCs, data of binary fraction and integrated spectral energy distribution}
\label{subsec:sample} 
In the test of the proposed method (Section 3), we choose twenty-one Galactic GCs, which have both binary fraction measurements based on the MSFL method and spectroscopic observations, as a test sample (referred to as Sample A).
In the application of the proposed method (Section 4), we use thirteen Galactic GCs, of which binary fractions have not been studied before, as a study sample (referred to as Sample B)

(1) {\sl Samples of GCs:} this work contains thirty-four GCs in the Milky Way (MW). 
(i) The main characteristics for the twenty-one GCs in Sample A are presented in Table~\ref{Tab:smpA-mc} (used in Section 3).
The first column is the name (including whether it is a PCC and/or AC GC?), the second to third columns are the corresponding references for the binary fraction data and for the integrated spectral energy distribution data (the number in the parentheses after 'S05' indicates the spectrum number by S05), 
the fourth to fifth columns are the number of horizontal branch (HB) stars, RR variables and blue stragglers (BSs) per luminosity $L_{\rm tot}$ in units of $10^4$ ($N_{\rm HB}$, $N_{\rm RR}$ and $N_{\rm BS}$),
and the sixth to fourteenth columns are the central concentration $c$ of the \citet{1962AJ.....67..471K} model, core radius $R_{\rm C}$, half mass radius $R_{\rm HM}$, absolute visual magnitude $M_{\rm V}$, metallicity [Fe/H], HB morphology HBR, distance from the Sun $R_{\rm sun}$ and distance from the Galactic center $R_{\rm GC}$.
The conclusions whether a GC is post core-collapsed are from the \citet{1996AJ....112.1487H} catalogue, 
those whether a GC is accreted from dSph galaxies are from \citet[][]{2010MNRAS.404.1203F, 2004MNRAS.355..504M} and \cite{2018ApJ...863L..28M}, 
$N_{\rm HB}$ and $N_{\rm RR}$ come from \citet[][]{2010A&A...517A..81G}, $N_{\rm BS}$ and $L_{\rm tot}$ are from \citet{2008A&A...483..183M}, 
and the data from the seven to fourteen columns are from the \citet{1996AJ....112.1487H} catalogue.
(ii) The main characteristics for the thirteen GCs in Sample B are presented in Table~\ref{Tab:smpB-mc} (used in Section 4), which is the same as Table~\ref{Tab:smpA-mc} except that the references for the binary fraction data are not given.

(2) {\sl Integrated spectral energy distribution of GCs:} the integrated spectral energy distributions of all thirty-four GCs are from P02, S05 and U17.

(3) {\sl Binary fraction of GCs:} the binary fractions of the twenty-one GCs in Sample A come from M12 and J15.
We use their binary-fractions with $q$>0.5, and the calibrations presented in Paper I are based on the binary-fractions with $q$>0.5.

\subsection{Calculations}
\label{subsec:calculations} 
To obtain the binary fraction of a given stellar cluster, we need to get SAFIs at 5\,\AA,  calibration coefficients, and the weights (see Eq.~\ref{Eq:method-1step4}),  which are introduced in the following sections.

\subsubsection{Transformation of spectrum to SAFIs at the resolution of 5\,\AA}
\label{subsec:transformation} 
Same as Paper I, we degrade the observed integrated spectral energy distributions to the resolution of 5\,\AA\ and transform them to SAFIs via the BC03 or our codes \citep{Bru03,Zha04}. 
Both codes use the SAFI definitions and directly integrate the flux within the index passband to obtain them. The local continuum is obtained by drawing a straight line from the midpoint of the blue pseudo- continuum to the midpoint of the red pseudo-continuum. For molecular indices, their values are further expressed in magnitudes.
In this work, we only use ten binary fraction-sensitive SAFIs, including five Balmer, two OIII, Ca4455, C$_2$4668 and TiO$_1$ indices.

\subsubsection{Updated calibration coefficients}
\label{subsec:updated} 
The transformation method of equivalent radius in Paper I has been introduced in Section 2.1.
In this work, we utilize the random scattering method and the surface density profile of the \citet{1962AJ.....67..471K} model to obtain equivalent radius by ensuring that the number of stars within the rectangle observed region equals to that within the equivalent region. 
In the top panels of Fig.~\ref{Fig:fig2}, we illustrate the difference in the equivalent radius between Paper I and this work, d$R_{\rm equ}$[Paper I$-$this work], as functions of the central concentration $c$ and core radius $R_{\rm C}$ for the forty-four GCs (which have both binary fraction and spectrum observations, used in Paper I) in the U17 and S05 spectrum samples.
From them, we see that the equivalent radii for the S05 observation in Paper I are smaller than in this work when $c$ is larger and $R_{\rm C}$ is smaller, because the slit height is much larger than the width.
In the bottom panels, we give the difference in the equivalent radius between the U17 and S05 observations, d$R_{\rm equ}$[U17$-$S05], as functions of $c$ and $R_{\rm C}$ for Paper-I and this work.
From them, we see that d$R_{\rm equ}$[U17$-$S05] in Paper I is similar to this work.

Using the equivalent radii in this work, we re-calculate the binary fractions within the equivalent regions and re-study the sensitivity of each SAFI to binary fraction and obtain a set of new calibration coefficients between binary fraction and SAFI variations at the resolution of 5\,\AA.
This set of coefficients is completely consistent with that in Paper I (see Table~\ref{Tab:slope}).
The main reason is that d$R_{\rm equ}$[U17$-$S05] in Paper I is similar to this work (see the bottom panels of Fig.~\ref{Fig:fig2}).

\subsubsection{Weights for SAFIs}
\label{subsec:weights} 
We obtain the binary fraction variation by weighting the binary fraction differences obtained via different SAFIs (see Eq.~\ref{Eq:method-1step1}). 
The sensitivity of each SAFI to the binary fraction is different. The SAFI range for all GCs represents the sensitivity of age and metallicity.
The weight for each SAFI is inversely proportional to the SAFI range for all GCs in Paper I, and it reflects the sensitivity of each SAFI to binary fraction.
In Table~\ref{Tab:weight}, we provide the weights for the ten binary fraction-sensitive SAFIs.

\section{Test of method}
\label{sec:result1}

\begin{figure}
\centering
\includegraphics[scale=0.4]{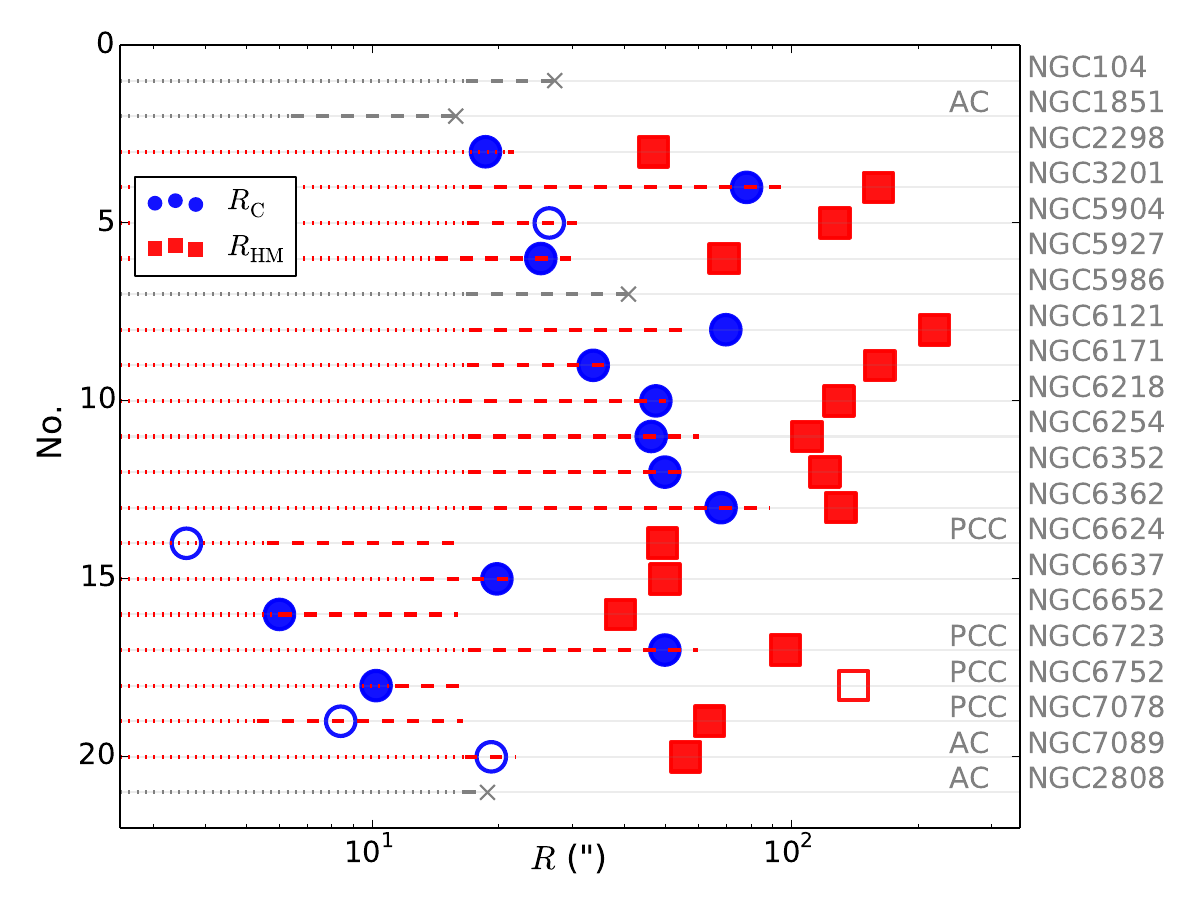} 
\caption{The maximal and minimal equivalent radii $R_{\rm equ,l}$ and $R_{\rm equ,u}$ (at both ends of dashed line) of the spectrum observed regions (corresponding to the rectangular regions in Fig.\,1), as well as the binary fraction photometric observation division radii $R_{\rm C}$ and $R_{\rm HM}$ (blue points and red squares, corresponding to the M1 and M2 regions in Fig.\,1) of M12.
If GC has neither $f_{(0-R_{\rm C})}$ nor $f_{(R_{\rm C}-R_{\rm HM})}$, we do not mark $R_{\rm C}$ and $R_{\rm HM}$, but instead mark the right end of the dashed line with a cross and use gray for the dotted and dashed lines. 
If the binary fraction within a region is provided by M12, the corresponding outer radius ($R_{\rm C}$ or $R_{\rm HM}$) is represented by a fill symbol.
The GC's PCC and AC attrbutes are labelled to the left of their name.
}
\label{Fig:requ-smpA}
\end{figure}

\begin{figure}
\centering
\includegraphics[scale=0.4]{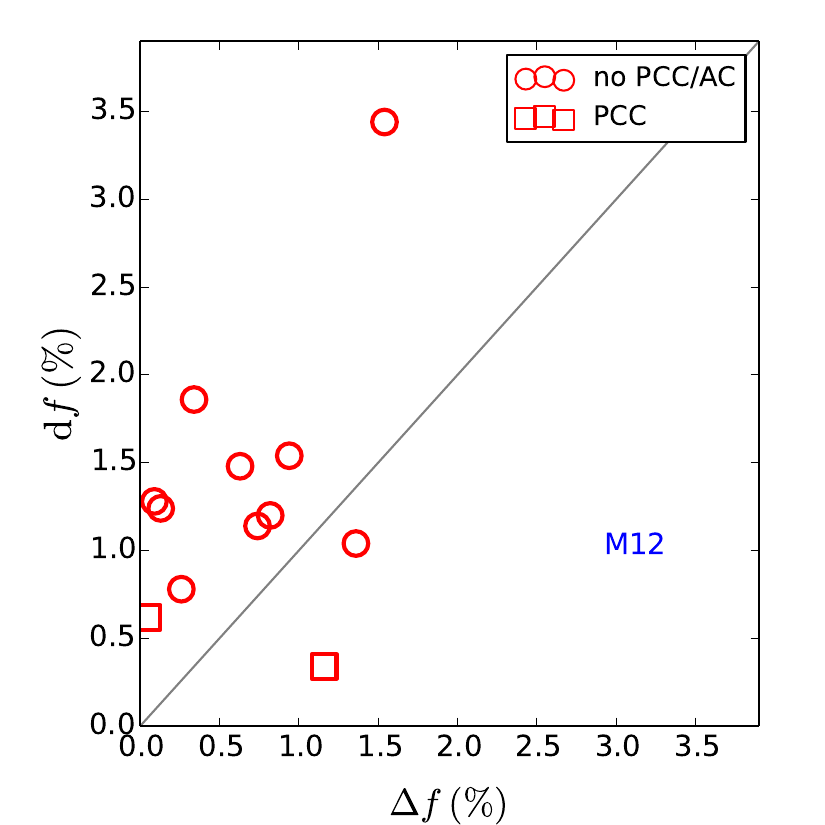} 
\caption{Comparison between the binary fraction variation $\Delta f$ in this work and d$f$(=0.2\,$f_{(0-R_{\rm C})}$) derived from the binary fraction data of M12 for the thirteen GC in Sample A.
}
\label{Fig:f-f-smpA}
\end{figure}

\begin{figure}
\centering
\includegraphics[scale=0.5]{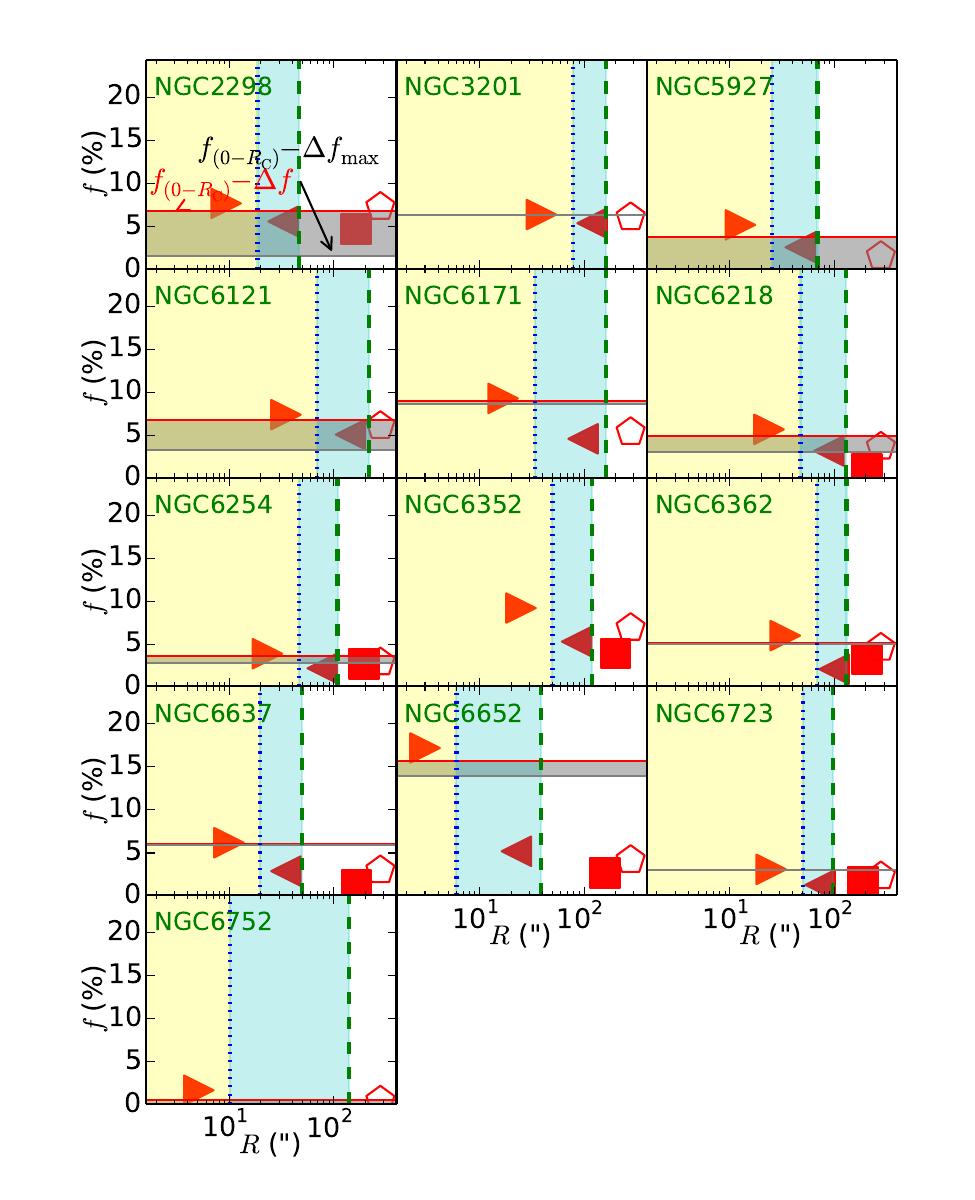} 
\caption{Comparison between the binary fraction in this work $f$ (=$f_{(0-R_{\rm C})}$$-$$\Delta$$f$,  gray horizontal line above) and the binary fractions (symbols) given by the MSFL method of M12 for the thirteen GCs in Sample A. Also shown are $f_{\rm min}$ (=$f_{(0-R_{\rm C})}$$-$$\Delta$$f_{\rm max}$, gray horizontal line below), $R_{\rm C}$ and $R_{\rm HM}$ (vertical dotted and dashed lines).
For M12, $f_{(0 - R_{\rm C})}$, $f_{(R_{\rm C} - R_{\rm HM})}$, $f_{(R_{\rm HM} - R_{\rm WFC})}$, and $f_{(0 - R_{\rm WFC})}$ are represented by solid right-triangles, left-triangles, rectangles and open stars (not for comparison), with each binary fraction given at the midpoint of the correspond region except for $f_{(0 - R_{\rm WFC})}$ (at $R_{\rm WFC}$).
}
\label{Fig:f-r-smpA}
\end{figure}

\begin{figure}
\includegraphics[scale=0.4]{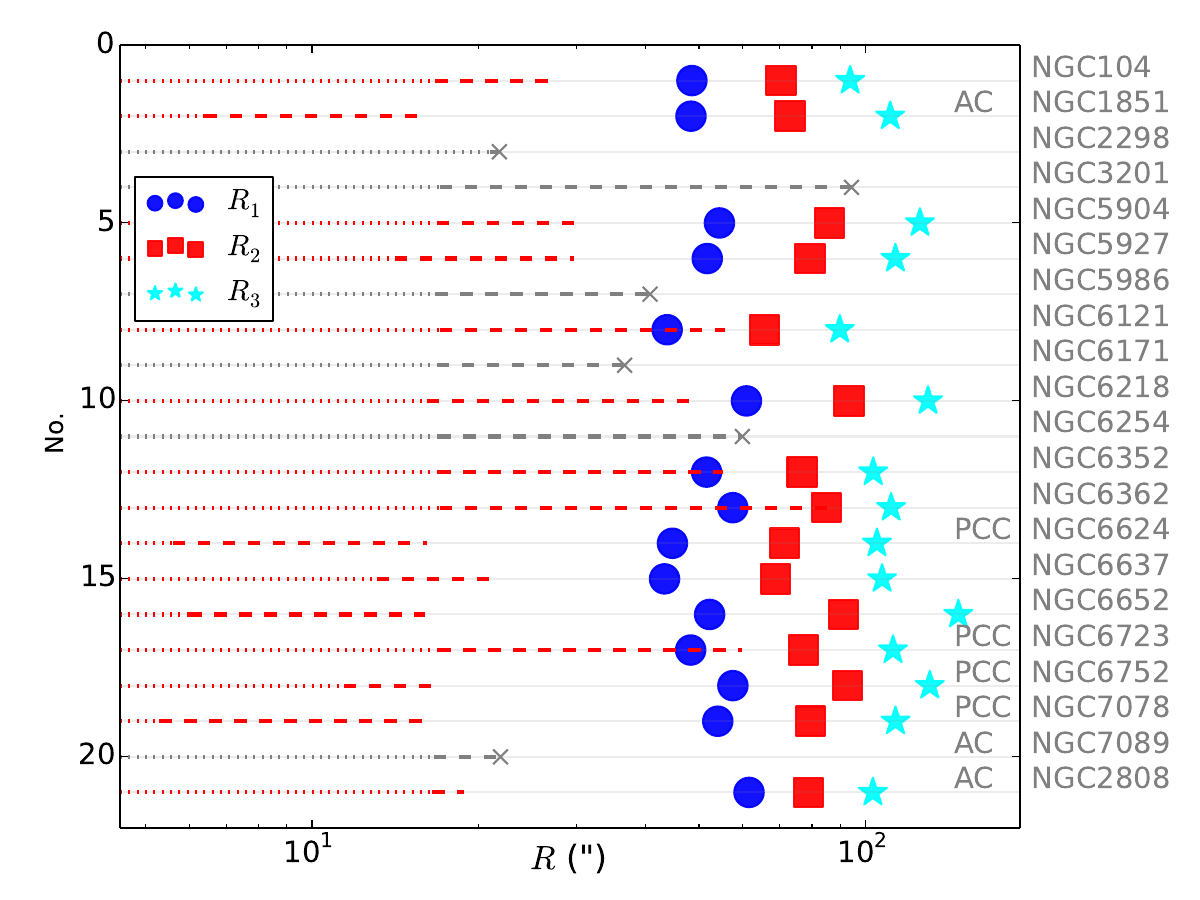} 
\caption{Similar to Fig.~\ref{Fig:requ-smpA}, the difference is that M12's binary fraction observation division radii $R_{\rm C}$ and $R_{\rm HM}$ are replaced by J15's division radii $R_{1}$, $R_{2}$ and $R_{3}$ (blue points, red squares, and cyan stars, corresponding to J1, J2 and J3 regions in Fig.~\ref{Fig:paperI}). 
}
\label{Fig:requ-smpA-J15}
\end{figure}

\begin{figure}
\centering
\includegraphics[scale=0.4]{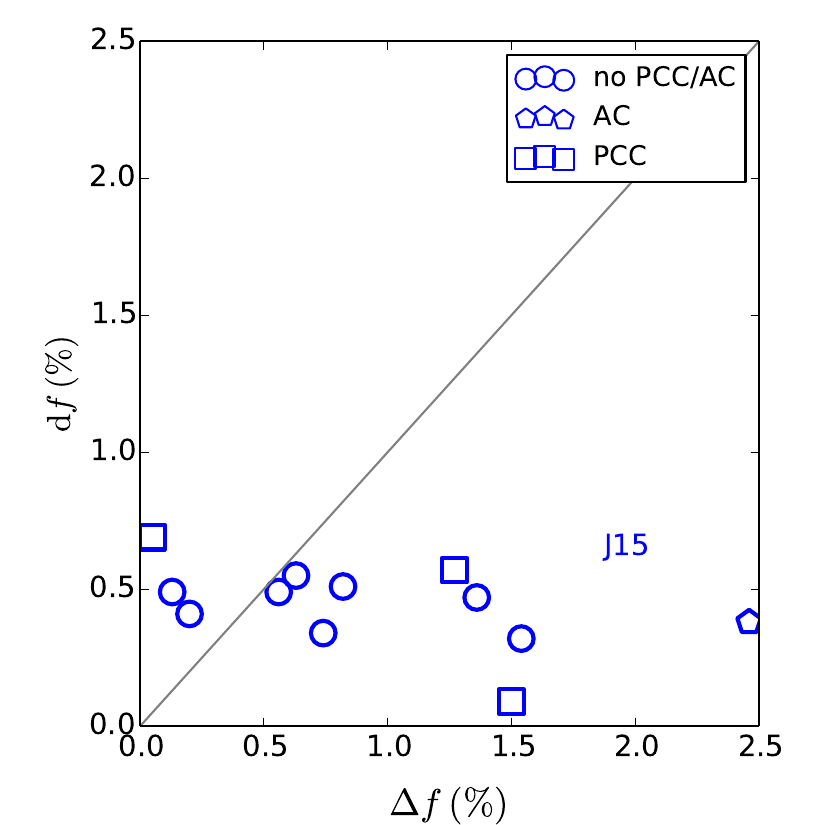} 
\caption{Similar to Fig.~\ref{Fig:f-f-smpA-J15}, but for d$f$(=0.1\,$f_{(0-R_1)}$) derived from the J15's binary fraction data.}
\label{Fig:f-f-smpA-J15}
\end{figure}

\begin{figure}
\includegraphics[width=8.8cm, height=10.cm]{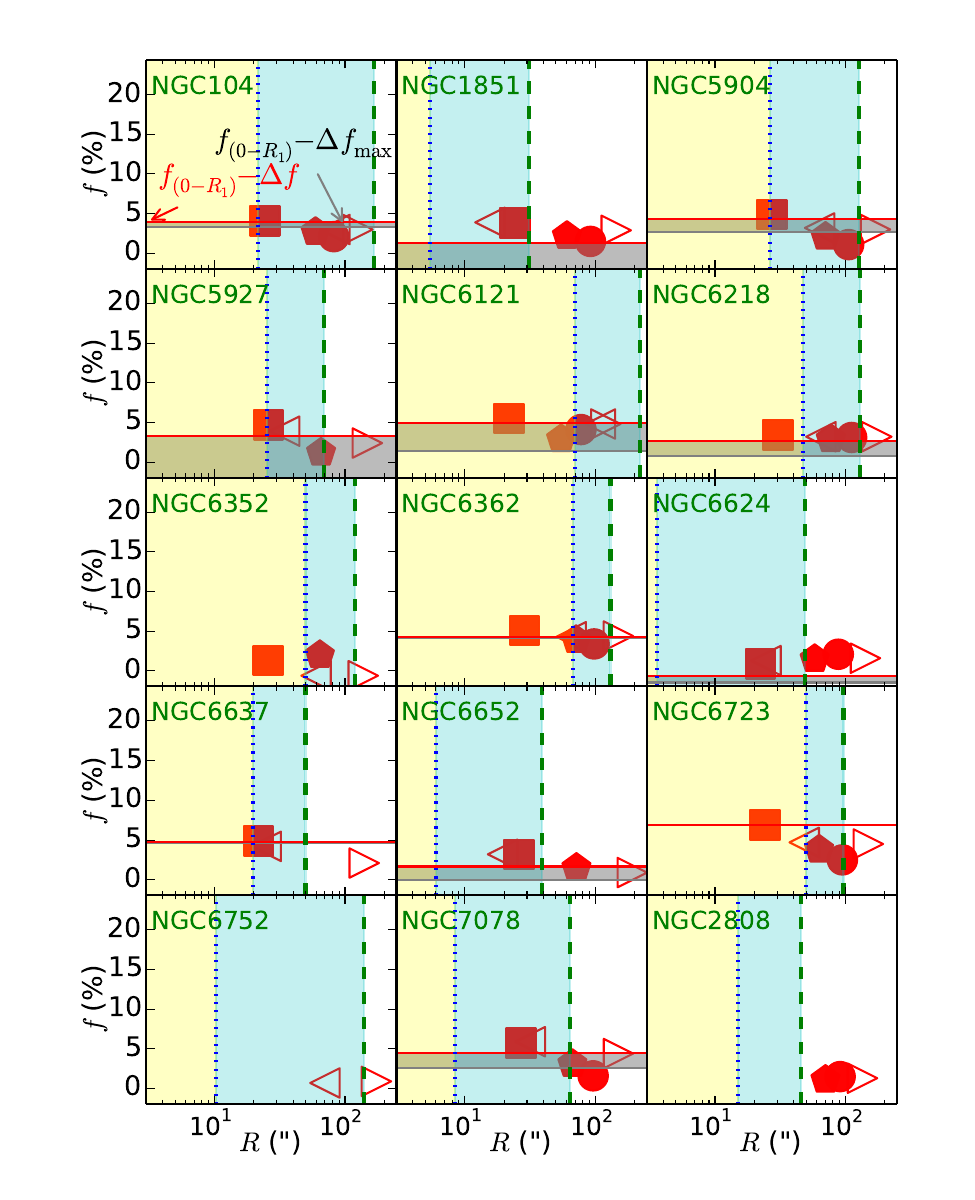} 
\caption{Similar to Fig.~\ref{Fig:f-r-smpA}, the differences are that there are fifteen GCs,  and the M12's binary fractions are replaced by J15's $f_{(0 - R_1)}$ (rectangles), $f_{(R_1 - R_2)}$ (red stars), $f_{(R_2 - R_3)}$ (red circles), $f_{(0 - R_{\rm HM})}$, and $f_{(0 - R_{\rm WFV})}$ (red open left- and right- triangles, not for comparison).}
\label{Fig:f-r-smpA-J15}
\end{figure}

\begin{figure*}
\includegraphics[width=15.5cm, height=8.50cm]{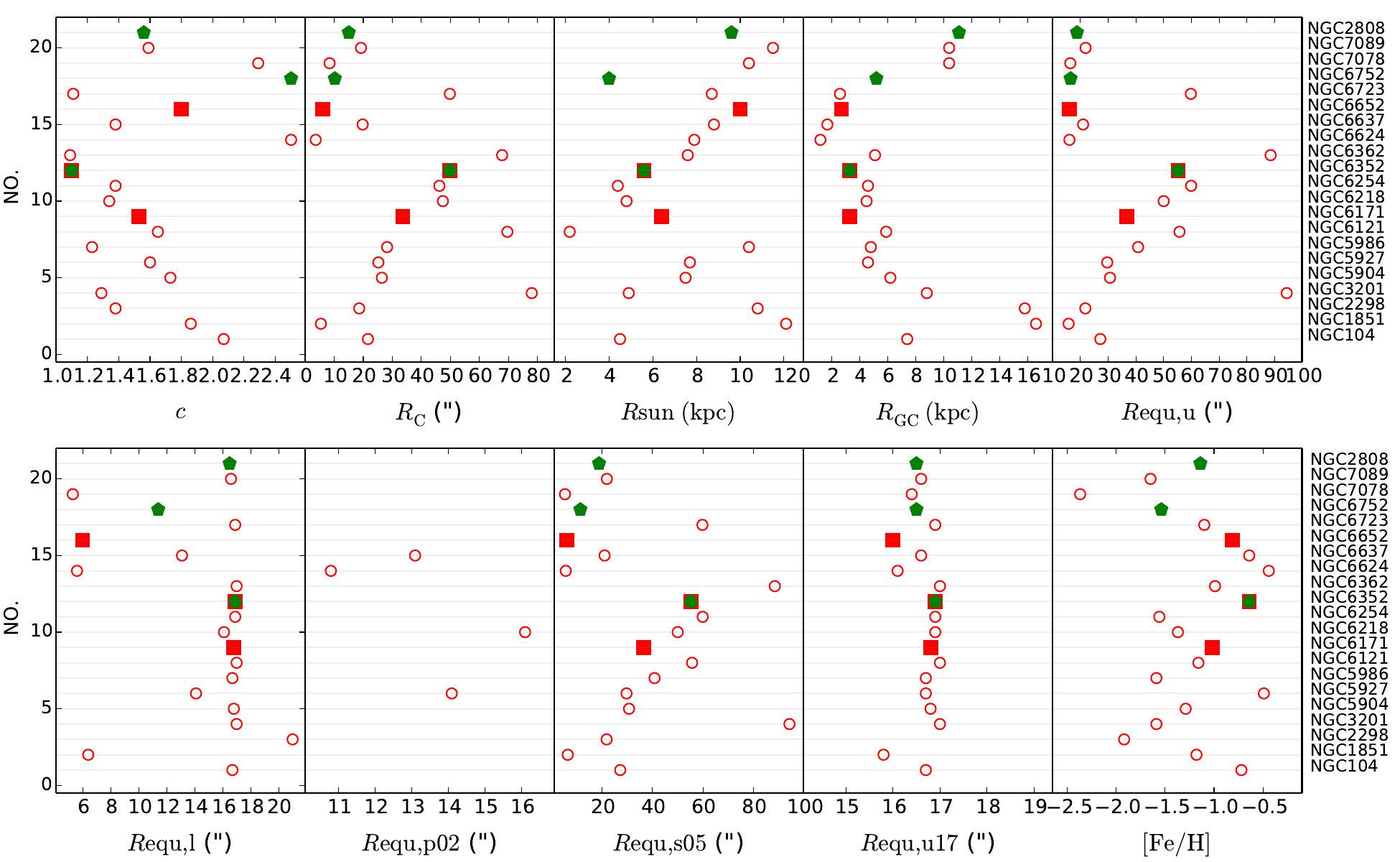} 
\caption{Distributions of various parameters (central concentration $c$, core radius $R_{\rm C}$, distance from the Sun $R_{\rm sun}$, distance from the Galactic center $R_{\rm GC}$, metallicity [Fe/H], the minimal equivalent radius $R_{\rm equ,l}$, the maximal equivalent radius $R_{\rm equ,u}$, the equivalent radii in the P02, S05 and U17 observations [$R_{\rm equ,p02}$, $R_{\rm equ,s05}$, $R_{\rm equ,u17}$]) for the twenty-one GCs in Sample A.
The GCs whose binary fractions do not match well with M12 and J15 are represented by red squares and green stars, respectively (for NGC6352, the red square is covered by a green star).
}
\label{Fig:pardis-smpA}
\end{figure*}

\begin{table}
\caption{The derive binary fraction variations $\Delta f_{\rm}$ and their maximal values $\Delta f_{\rm max}$ for the twenty-one GCs in Sample A.}
\centering
\begin{tabular}{llrr}
\hline
 ID & Name  & $\Delta f_{\rm}$\,(\%)  & $\Delta f_{\rm max}$\,(\%)\\
\hline
 1 & NGC104 & 0.20 & 0.78 \\
 2 &NGC1851 & 2.46 & 22.84 \\
 3 &NGC2298 & 0.94 & 6.08 \\
 4 &NGC3201 & 0.09 & 0.09 \\
 5 &NGC5904 & 0.56 & 2.20 \\
 6 &NGC5927 & 1.36 & 8.51 \\
 7 &NGC5986 & 0.22 & 0.88 \\
 8 &NGC6121 & 0.63 & 4.13 \\
 9 &NGC6171 & 0.34 & 0.66 \\
 10 &NGC6218 & 0.74 & 2.59 \\
 11 &NGC6254 & 0.26 & 1.13 \\
 12 &NGC6352 & 10.05 & 10.05 \\
 13 &NGC6362 & 0.82 & 0.97 \\
 14 &NGC6624 & 1.50 & 2.36 \\
 15 &NGC6637 & 0.13 & 0.25 \\
 16 &NGC6652 & 1.54 & 3.21 \\
 17 &NGC6723 & 0.05 & 0.05 \\
 18 &NGC6752 & 1.16 & 4.23 \\
 19 &NGC7078 & 1.27 & 3.12 \\
 20 &NGC7089 & 0.22 & 1.07 \\
 21 &NGC2808 & 0.55 & 1.43 \\
\hline
\end{tabular}\\
\label{Tab:drfb-smpA}
\end{table}

Up to now, there are no such stellar systems which have a suitable set of integrated spectra (as suggested by us in Fig.~\ref{Fig:method}), so we currently cannot verify every step of the proposed method. 
Some Galactic GCs have several integrated spectral energy distributions within different regions and binary fraction measurements, so we can verify one step of the proposed method. 
Sample A is composed of these GCs.

We first obtain the GCs' binary fraction variations in Section~\ref{subsec:variationsA}, compare binary fraction variations with those derived from the binary fraction data, and compare binary fractions with those obtained by the MSFL method to testify one of steps in the method in Section 3.2.
This test can demonstrate whether the proposed binary fraction determination method is feasible.
Moreover,  we analyze the reasons for the differences in Section 3.3 and give some suggestions for spectroscopic and photometric observations in order to get binary fraction and more accurate recalibration relation in Section 3.4.

\subsection{Binary fraction variations for the GCs in Sample A}
\label{subsec:variationsA}
In the calculations of binary fraction variations, we use the fitted SAFIs (at the resolution of 5\,\AA, Section~\ref{subsec:transformation}) within the minimal and maximal equivalent regions for each GC since each observed spectrum has observation error.
By combining with the calibrations between binary fraction and SAFI variations (Eq.~\ref{Eq:calib} and Table~\ref{Tab:slope}, Section~\ref{subsec:updated}) and weights for the ten binary fraction sensitive SAFIs (Table~\ref{Tab:weight}, Section~\ref{subsec:weights}), we use Eq.~\ref{Eq:method-1step4} to calculate the binary fraction variations $\Delta$$f$ and their maximal values  $\Delta$$f_{\rm max}$ (the maximum for the set of binary fraction variations \{$\Delta$$f_k$\} from the ten binary fraction sensitive SAFIs) for the twenty-one GCs in Sample A.The results are presented in Table~\ref{Tab:drfb-smpA}.

\subsection{Comparison with the M12 and J15 results}
\label{subsec:Comparison} 

\subsubsection{Comparison with the M12 results}
\label{subsec:ComparisonM12} 
What we presented in Table~\ref{Tab:drfb-smpA} is the binary fraction variation $\Delta f$ between the two observed regions. M12 provided the binary fractions $f$ within the ranges of 0--$R_ {\rm C}$, $R_ {\rm C}$--$R_ {\rm HM}$, $R_ {\rm HM}$--$ R_ {\rm WFC}$ (HST's ACS/WFC radius) and 0--$ R_ {\rm WFC}$. We used two comparison methods: one is to compare the binary fraction variation in this work $\Delta f$ with d$f$ derived from the binary fraction data of M12, and the other is to compare the binary fraction $f$ with those from the MSFL method.

In the first comparison method, we not only need to know the sizes, positions and the number of stars (see Eq.~\ref{Eq:f2-1}) for the spectrum observed regions (corresponding to the rectangular regions in Fig.~\ref{Fig:paperI}), but also need to know the sizes and locations of the binary fraction observed regions (M1, M2, J1, ..., J5 regions in Fig.~\ref{Fig:paperI}), and whether the binary fraction is given in the corresponding region.
The observed regions of M12 are  bounded by core radius $R_ {\rm C}$ and half mass radius $R_ {\rm HM}$.
In Fig.~\ref{Fig:requ-smpA}, we present the maximal and minimal equivalent radii $R_{\rm equ,l}$ and $R_{\rm equ,u}$ for the spectrum observed regions of each GC, as well as the M12's binary fraction photometric observation division radii $R_ {\rm C}$ and half mass radius $R_ {\rm HM}$.
We want to use $f_{(0 - R_{\rm C})}$ to represent the binary fraction variation d$f$, and there are thirteen GCs with $f_{(0 - R_{\rm C})}$ in Sample A. 
For these thirteen GCs, from this plot we can see that (i) the minimal equivalent radii $R_{\rm equ,l}$<$R_{\rm C}$ for ten GCs and $R_{\rm equ,l}$$\simeq$$R_{\rm C}$ for three GCs (NGC2298, NGC6652, NGC6752), (ii) the maximal equivalent radii $R_{\rm equ,u}$$\simeq$$R_{\rm C}$ for eleven GCs and $R_{\rm C}$<$R_{\rm equ,u}$<$R_{\rm HM}$ for two GCs (NGC6652, NGC6752). 
For these thirteen GCs, we approximate the binary fraction within the minimal equivalent region as $f_{\rm 1}$$\simeq$$f_{(0 - R_{\rm C})}$. In fact, for the ten GCs with $R_{\rm equ,l}$<$R_{\rm C}$, $f_{\rm 1}$ will be slightly higher than $f_{(0 - R_{\rm C})}$ because the steep surface density profile leads to the average binary fraction being dominated by the value within the central region.
The binary fraction within the maximal equivalent region $f_2$ is obtained by
\begin{equation}
f_2 =
\begin{cases}
f_{(0 - R_{\rm C})}   & \text{if }  R_{\rm equ,u} \leq R_{\rm C}   \\
\frac{N_{\rm OC}\ f_{(0 - R_{\rm C})}+N_{\rm CU}\ f_{(R_{\rm C} - R_{\rm equ,u})}}{N_{\rm OC}+N_{\rm CU}} & \text{if }  R_{\rm C}<R_{\rm equ,u}<R_{\rm HM}, \\
\end{cases}
\label{Eq:f2-1}
\end{equation}
in which $f_{(0 - R_{\rm C})}$ and $f_{(R_{\rm C} - R_{\rm equ,u})}$ are the binary fractions, $N_{\rm OC}$ and $N_{\rm CU}$ are the number of stars within the ranges of 0--$R_{\rm C}$ and $R_{\rm C}$--$R_{\rm equ,u}$.  
$N_{\rm OC}$ and $N_{\rm CU}$ can be obtained by the \citet{1962AJ.....67..471K} surface density profile (Eq.~\ref{Eq:psd}) and the equivalent radii shown in Fig.~\ref{Fig:requ-smpA}. 
Assuming that binary fraction is constant within the range of $R_{\rm C}$--$R_{\rm HM}$, then $f_{(R_{\rm C} - R_{\rm equ,u})}$=$f_{(R_{\rm C} - R_{\rm HM})}$.
The calculation results indicate that $f_{\rm 2}$ is in the range of 0.80--1.00$f_{(0 - R_{\rm C})}$ (0.67$f_{(0 - R_{\rm C})}$ for NGC6752 by assuming $f_{(R_{\rm C} - R_{\rm HM})}$=0, M12 did not provide the binary fraction within the range of $R_{\rm C}$--$R_{\rm HM}$ for this GC).
In fact, the binary fraction within the range of $R_{\rm C}$--$R_{\rm equ,u}$ is larger than the average $f_{(R_{\rm C} - R_{\rm HM})}$, so the second term of the numerator in the second case of Eq.~\ref{Eq:f2} and $f_{\rm 2}$ will be larger.
In summary, the difference d$f$(=$f_{\rm 1}$$-$$f_2$) varies from 0 to 0.2$f_{(0-R_{\rm C})}$ (d$f$ will tend towards 0 in the cases of $R_{\rm equ,u} $$\simeq$$ R_{\rm C}$, very steep [few stars in the range of $R_{\rm C}$--$R_{\rm equ,u}$] or very flat binary fraction profiles).

In Fig.~\ref{Fig:f-f-smpA}, we compare the binary fraction variations in this work $\Delta f$ (Table~\ref{Tab:drfb-smpA}) with d$f$ (=0.2$f_{(0 - R_{\rm C})}$) derived from the M12's binary fraction data for the thirteen GCs. From it, we can see that the binary fraction variations in this work are consistent with 0.2$f_{(0 - R_{\rm C})}$.

Moreover, in Fig.~\ref{Fig:f-r-smpA}, we compare the binary fractions in this work $f$ (=$f_{(0 - R_{\rm C})}$-$\Delta$$f$) with those from the M12's MSFL method for the thirteen GCs.
For these thirteen GCs, the binary fraction within the range of 0--$R_{\rm equ,u}$ is $f$$\simeq$$f_{(0 - R_{\rm C})}$$-$$\Delta f$ because $f_{\rm 1}$$\simeq$$f_{(0 - R_{\rm C})}$. 
Therefore, in Fig.~\ref{Fig:f-r-smpA}, we present $f$, $f_{\rm min}$ (=$f_{(0 - R_{\rm C})}$$-$$\Delta f_{\rm max}$), M12's $f_{(0 - R_{\rm C})}$, $f_{(R_ {\rm C} - R_ {\rm HM})}$, $f_{(R_ {\rm C} - R_ {\rm WFC})}$ and $f_{(0 - R_ {\rm WFC})}$ (not for comparison) for the thirteen GCs.
The maximal  equivalent radius $R_{\rm equ}$ is approximately equal to $R_{\rm C}$ for twelve GCs, and $f$ should be compared with $f_{(0-R_{\rm C})}$. For NGC6752, $R_{\rm C}$< $R_{\rm equ}$<$R_{\rm HM}$, $f$ should be between $f_{(0-R_{\rm C})}$ and $f_{(0-R_{\rm HM})}$ (see Fig.~\ref{Fig:requ-smpA}).
From Fig.~\ref{Fig:f-r-smpA}, we can see that the binary fractions are consistent with those of M12 for all GCs (except NGC6171, NGC6352, and NGC6652, for which we will analyze the reasons in Section~\ref{subsec:reasons}).

\subsubsection{Comparison with the J15 results}
\label{subsec:ComparisonJ15} 
For the same reason as the second paragraph of Section~\ref{subsec:ComparisonM12}, in Fig.~\ref{Fig:requ-smpA-J15}, we present the maximal and minimal equivalent radii $R_{\rm equ,l}$ and $R_{\rm equ,u}$ for each GC, as well as the first, second, and third division radii for binary fraction photometric observations  of J15 (corresponding to the J1, J2 and J3 regions in Fig.~\ref{Fig:paperI}). 
We want to use $f_{(0-R_1)}$ to represent the binary fraction variation d$f$, where fifteen GCs in Sample A have $f_{(0-R_1)}$.
For these fifteen GCs, we can see from Fig.~\ref{Fig:requ-smpA-J15} that (i) the minimal radii $R_{\rm equ,l}$<$R_1$, and (ii) the maximal radii $R_{\rm equ,u}$<$R_1$ for twelve GCs and $R_1$<$R_{\rm equ,u}$<$R_2$ for three GCs (NGC6121, NGC6362, NGC6723).
We approximate the binary fraction within the minimal equivalent range as $f_1$$\simeq$ $f_{(0 - R_1)}$, but the actual $f_1$ will be lager for these fifteen GCs. The binary fraction within the maximal equivalent region $f_2$ is obtained by
\begin{equation}
f_2 =
\begin{cases}
f_{(0 - R_{\rm 1})}   & \text{if }  R_{\rm equ,u} \leq R_{\rm 1}   \\
\frac{N_{\rm O1}\ f_{(0 - R_{\rm 1})}+N_{\rm 1U}\ f_{(R_{\rm 1} - R_{\rm equ,u})}}{N_{\rm O1}+N_{\rm 1U}} & \text{if }  R_{\rm 1}<R_{\rm equ,u}<R_{\rm 2}, \\
\end{cases}
\label{Eq:f2}
\end{equation}
in which $f_{(0 - R_{\rm 1})}$ and $f_{(R_{\rm 1} - R_{\rm equ,u})}$ are the binary fractions, $N_{\rm O1}$ and $N_{\rm 1U}$ are the number of stars within the ranges of 0--$R_{\rm 1}$ and $R_{\rm 1}$--$R_{\rm equ,u}$.
Assuming that the binary fraction in the range of ${R_{\rm 1} - R_{\rm equ,u}}$, $f_{(R_{\rm 1} - R_{\rm equ,u})}$=$f_{(R_{\rm 1} - R_{\rm 2})}$, we calculate that $f$ is in the range of 0.9--1.0$f_{(0 - R_1)}$. 
Therefore, d$f$ (=$f_2$$-$$f_1$) is in the range of 0.0--0.1$f_{(0 - R_1)}$.

In Fig.~\ref{Fig:f-f-smpA-J15}, we compare the binary fraction variations in this work $\Delta f$ with d$f$ (=0.1$f_{(0 - R_1)}$) derived from the J15's binary fraction data for the fifteen GCs in Sample A. From it, we can see that binary fraction variations in this work are consistent with 0.1$f_{(0 - R_1)}$.

In Fig.~\ref{Fig:f-r-smpA-J15}, we compare the binary fraction in this work $f$(=$f_{(0 - R_1)}$$-$$\Delta f$) with J15's  $f_{(0 - R_1)}$, $f_{(R_ {\rm 1} - R_ {\rm 2})}$, and $f_{(R_ {\rm 2} - R_ {\rm 3})}$ for the fifteen GCs.
The maximal equivalent radius $R_{\rm equ}$ is comparable with $R_{\rm 1}$ for twelve GCs, $f$ should be compared with $f_{(0-R_{\rm 1})}$. For NGC6121, NGC6362 and NGC6723, $R_{\rm 1}$< $R_{\rm equ}$<$R_{\rm 2}$, $f$ should be between $f_{(0-R_{\rm 1})}$ and $f_{(R_{\rm 1}-R_{\rm 2})}$ (see Fig.~\ref{Fig:requ-smpA-J15}).
From Fig.~\ref{Fig:f-r-smpA-J15}, we can see that binary fractions in this work are in good agreement with those of J15 for all GCs (except NGC6352, NGC6752, and NGC2808, for which we will analyze the reasons in Section~\ref{subsec:reasons}).

\subsubsection{Summary for the comparisons}
\label{subsec:ComparisonSum} 
In this part, we summarize the comparison results in the binary fraction between this work and previous work (M12 and J15) for the twenty-one GCs in Sample A (Sections 3.2.1 and 3.2.2). The binary fractions match well with those of M12 or J15 for sixteen GCs. 
For NGC6171, the derived binary fraction does not match well with that of M12, and coincidentally, J15 did not provide the binary fraction for this GC. 
For NGC2808, its binary fraction does not match well with that of J15, and M12 did not have its binary fraction (only had $f_{(0 - R_{\rm WFC})}$).
For NGC6352, its binary fraction does not match well with both M12 and J15 ($\Delta f$=10\% for NGC6352). 
For NGC5986 and NGC7089, M12 and J15 did not have their binary fractions (M12 had $f_{(0 - R_{\rm WFC})}$ and J15 had $f_{(0 - R_{\rm HM})}$).

\subsection{Reasons for the differences}
\label{subsec:reasons} 
In order to analyze the reasons for the differences, in Fig.~\ref{Fig:pardis-smpA} we plot the distributions of the central concentration $c$, core radius $R_{\rm C}$, distance from the Sun $R_{\rm sun}$, distance from the Galactic center $R_{\rm GC}$, maximal equivalent radius $R_{\rm equ, u}$, minimal equivalent radius $R_{\rm equ, l}$, equivalent radii in the P02, S05 and U17 observations ($R_{\rm equ, S05}$, $R_{\rm equ, P02}$ and $R_{\rm equ, U17}$) for the twenty-one GCs in Sample A (see Table~\ref{Tab:smpA-mc}).

First, we analyze the three GCs whose binary fractions do not agree well with those of M12 (red solid squares in Fig.~\ref{Fig:pardis-smpA}). 
For NGC6171, the discrepancy is likely caused by spectroscopic or/and photometric (used to the binary fraction determination) observation errors. 
For NGC6352, the difference is most likely due to spectroscopic observation error because its binary fraction does not match well with both M12 and J15. 
For NGC6652, the difference maybe is caused by photometric observation errors or by its relative small $R_{\rm C}$, $R_{\rm equ, l}$ and $R_{\rm equ, u}$ (see the top-second, top-fifth, and bottom-left panels of Fig.~\ref{Fig:pardis-smpA}).

Next, we discuss the three GCs whose binary fractions differ from those of J15 (green solid stars in Fig.~\ref{Fig:pardis-smpA}). 
NGC6352 has already been discussed above. 
For NGC6752 and NGC2808, $f_{(0 - R_1)}$ by J15 is negative. 
In addition, for NGC6752, the difference maybe is likely due to its very large $c$ and very low $R_{\rm C}$ (see the top-first and top-second panels of Fig.~\ref{Fig:pardis-smpA}).
For NGC2808, the difference is perhaps due to a very small $R_{\rm C}$ (see the top-left panel of Fig.~\ref{Fig:pardis-smpA}).

\subsection{Suggestions for the observations of the binary fraction determination and $\Delta$$f$ calibrations from $\Delta$SAFIs}
\label{subsec:suggestions} 
From the above analysis, it can be seen that we can derive the binary fraction variation of a stellar system. Through multiple identical steps, we can obtain the binary fraction of the whole star system.

However, if we want to determine the binary fraction of a stellar system, we need to obtain its integrated spectral energy distributions within different regions, whose positions and sizes must be suitable. 
For example, the equivalent radius difference of the selected nuclear regions should not be too large, and the difference can be appropriately increased when the regions are more than half mass radius $R_{\rm HM}$ from the center of a stellar system.

In addition, if we want to determine the binary fraction of a stellar system, we also need accurate $\Delta$$f$ calibrations in terms of $\Delta$SAFIs. 
Therefore, we need to obtain the binary fractions within the appropriate regions, establish a more accurate binary fraction profile, and thus obtain a set of more accurate $\Delta$$f$$-$$\Delta$SAFI calibration coefficients.

\section{Results and discussions}
\label{sec:result2}
\begin{figure}
\includegraphics[scale=0.4]{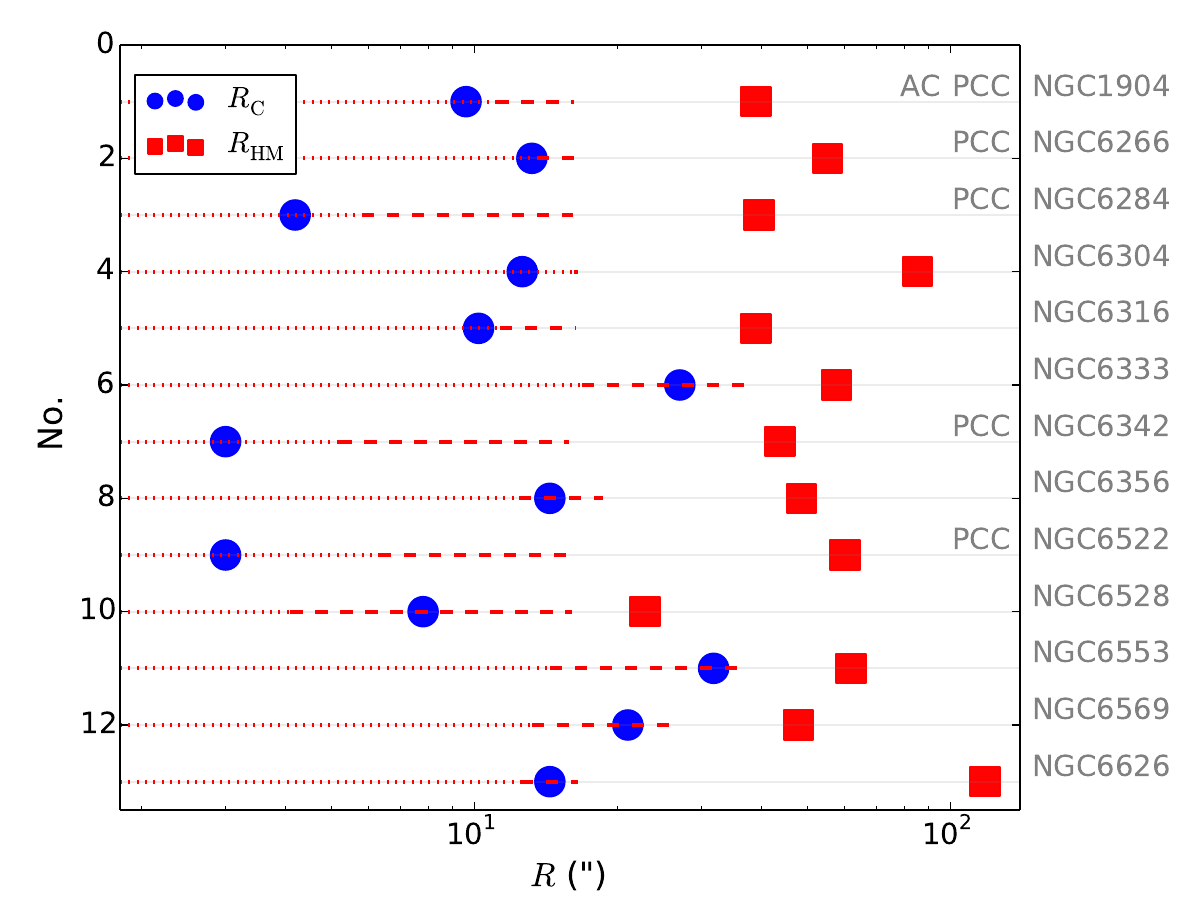} 
\caption{Minimal and maximal equivalent radii $R_{\rm equ,l}$ and $R_{\rm equ,u}$ (at both ends of  dashed lines) in the spectroscopic observations, core radius $R_{\rm C}$ (blue squares) and half mass radius $R_{\rm HM}$ (red rectangles) for the thirteen GCs in Sample B. Its PCC and ACC properties are labelled to the left of the GC's name.
}
\label{Fig:requ-smpB}
\end{figure}

%
\begin{figure*}
\includegraphics[scale=0.5]{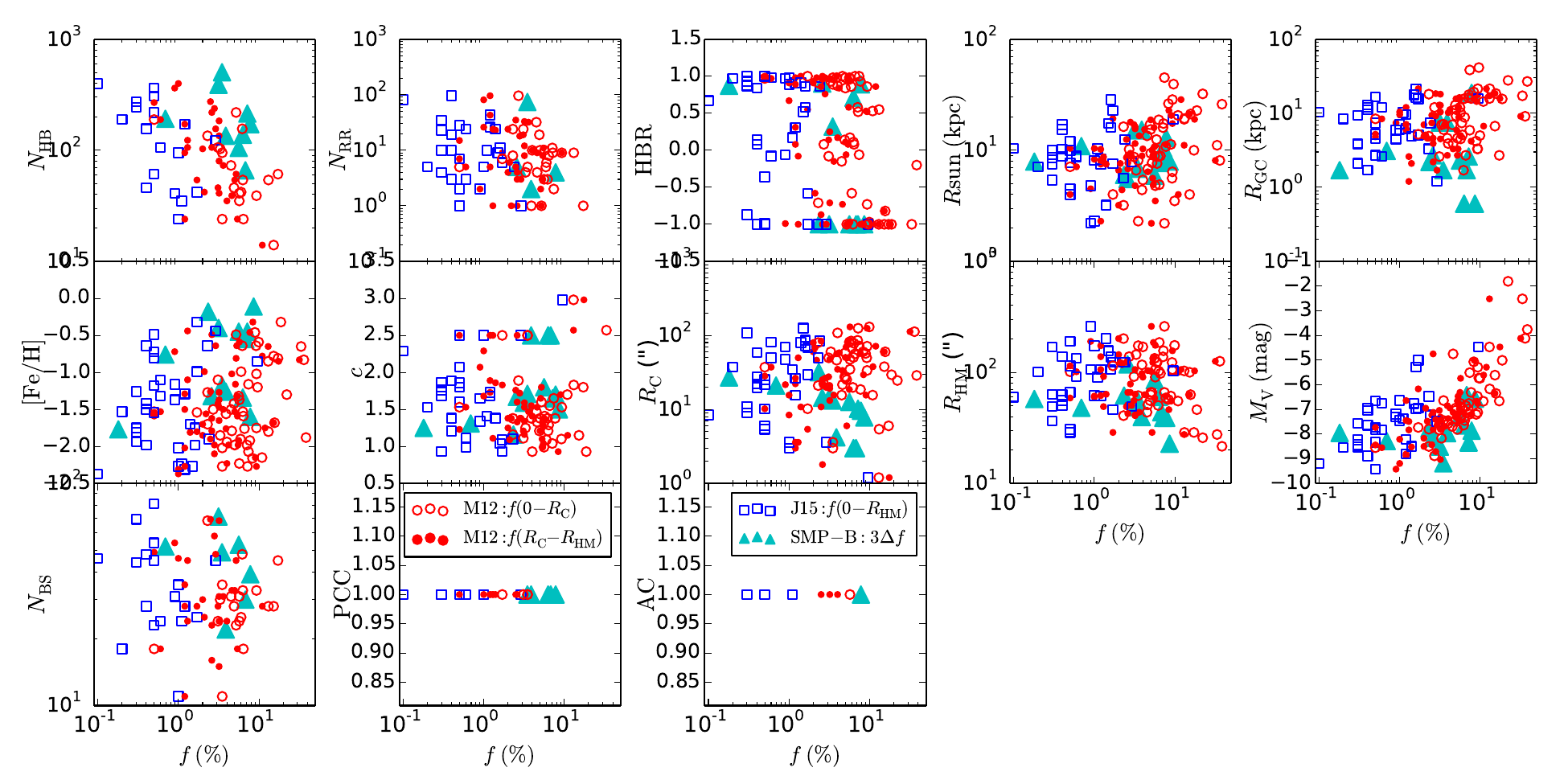} 
\caption{Relations between binary fractions $f$ and various parameters (number of HB stars $N_{\rm HB}$, number of RR variables $N_{\rm RR}$, HB morphology HBR, distance for the Sun $R_{\rm sun}$, distance from the Galactic center $R_{\rm GC}$, metallicity [Fe/H], central concentration $c$, core radius $R_{\rm C}$,  half mass radius $R_{\rm HM}$, absolute visual magnitude $M_{\rm V}$, number of BSs $N_{\rm BS}$) and processes (PCC and AC).
The thirteen GCs in Sample B  are in cyan solid triangles, the GCs with $f_{(0 - R_{\rm C})}$ and $f_{(R_{\rm C}-R_{\rm HM})}$ in the M12 study are in red open and solid circles, and those GCs with $f_{(0 - R_{\rm 1})}$ in the J15 study are in blue open squares.
}
\label{Fig:f-par}
\end{figure*}

\begin{table}
\caption{The binary fraction variations $\Delta f$ and their maximal values $\Delta f_{\rm max}$ for the thirteen GCs in Sample B.}
\begin{tabular}{llrr}
\hline
ID & Name & $\Delta f$\,(\%) & $\Delta f_{\rm max}$\,(\%)  \\
\hline
  1   &   NGC1904  &  2.60 & 11.55      \\
  2   &   NGC6266  &  1.16 &  2.98       \\
  3   &   NGC6284  &  1.29 &  3.05  \\
  4   &   NGC6304  &  1.87 &  3.88       \\
  5   &   NGC6316  &  2.39 & 24.27      \\
  6   &   NGC6333  &  0.06 &  0.06        \\
  7   &   NGC6342  &  2.26 & 20.15    \\
  8   &   NGC6356  &  1.05 &  6.99        \\
  9   &   NGC6522  &  2.10 &  6.47     \\
 10  &   NGC6528  &  2.86 & 46.62   \\
 11  &   NGC6553  &  0.78 &  3.33         \\
 12  &   NGC6569  &  0.23 &  0.73        \\
 13  &   NGC6626  &  0.86 & 11.38       \\
\hline
\end{tabular}
\label{Tab:drfb-smpB}
\end{table}

\subsection{Binary fractions for the thirteen GCs in Sample B}
\label{subsec:41} 
In this section, we use the thirteen GCs in Sample B. The binary fractions of the GCs in Sample B have not been studied before. 
Overall, the thirteen GCs used in this section are those that are far away from us and have brighter absolute visual magnitude. Interestingly, two GCs are associated with binary millisecond pulsar (BMP), one is a PCC and AC GC (from Sgr dSph galaxy), and four are PCC GCs.

For the SAFIs at the resolution of 5\,\AA\,(Section~\ref{subsec:transformation}), we also use the fitted SAFIs within the minimal and maximal equivalent regions for each GC. By combining with the update binary fraction variation calibrations in terms of $\Delta$SAFIs (Section~\ref{subsec:updated}) and weights for the ten binary fraction-sensitive SAFIs (Section~\ref{subsec:weights}), we calculate the binary fraction variations.

In Table~\ref{Tab:drfb-smpB}, we present the binary fraction variations for the thirteen GCs in Sample B. In order to discuss the relationships between binary fraction and various parameters, we need to know the locations and sizes of the spectrum observed regions in order to convert $\Delta$$f$ into binary fraction. 
In Fig.~\ref{Fig:requ-smpB}, we give the minimal equivalent radius $R_{\rm equ,l}$,  the maximal equivalent radius $R_{\rm equ,u}$, core radius $R_{\rm C}$ and half mass radius $R_{\rm HM}$ for these GCs.
From Fig.~\ref{Fig:requ-smpB}, we can see that $R_{\rm equ,l}$$\simeq$$R_{\rm C}$ for ten GCs, $R_{\rm equ,l}$<$R_{\rm C}$ for three GCs, $R_{\rm C}$<$R_{\rm equ,u}$<$R_{\rm HM}$ for all GCs.
The ratios of $R_{\rm equ,l}$ and $R_{\rm equ,u}$ to $R_{\rm C}$ and $R_{\rm HM}$ are similar to those of GCs in Sample A.
In Section 3.2.1, we conclude that d$f$=0-0.2$f_{(0-R_{\rm C})}$, and $f_2$ and d$f$ will be larger.
Therefore, we use the relation of d$f$$\simeq$0.3$f_{(0 - R_{\rm C})}$ to convert $\Delta f$ to binary fraction.

We first discuss the binary fractions for the PCC, AC, BMP-related GCs, and then those for the other GCs.
NGC1904 is a PCC and AC GC, its binary fraction variation is about 2.6\% and binary fraction is $f$$\simeq$7.8\%.
\citet{2018Ap&SS.363...97L} have reached a conclusion of $f$<20\%.
By comparing with the binary fractions of PCC GCs by M12 ($f_{(0 - R_{\rm C})}$<4\%, see the bottom-second panel of Fig.~\ref{Fig:f-par}), we find that the binary fraction obtained in this work is relative large.
The binary fractions of AC GCs may contain the MW's accretion information, but this goes beyond the scope of this study. 
NGC6266 contains the second most BMPs among all GCs and its binary fraction in this work is $f$$\simeq$3.5\%.
NGC6316 has similar color-magnitude diagram to 47 Tuc \citep {2023ApJ...942..104D}, which has the third most BMPs ($f$$\simeq$0.7\% in this work,  0.9\% for WFC by M12), and its binary fraction is $f$$\simeq$7.2\%. 
The binary fractions of these BMP-related GCs contribute to understanding the formation of BMPs and their physical processes.
The other ten GCs are non-PCC and non-AC GCs and their binary fractions can reach $f$$ \simeq$8.7\%.

\subsection{Relations between binary fraction and various parameters and processes}
\label{subsec:41} 
M12 has discussed the correlations between binary fraction and many parameters, and concluded that binary fraction is related to the number of BSs (mainly for $f_{(0-R_{\rm C})}$) and absolute visual magnitude. 
By combining the binary fractions of the thirteen GCs in this work with $f_{(0-R_{\rm C})}$ and $f_{(R_{\rm C}-R_{\rm HM})}$ of M12 and $f_{(0-R_{\rm 1})}$ of J15, we analyze the relationships between binary fraction and the number of HB stars, RR variables and BSs per luminosity $L_{\rm tot}$ in units of $10^4$ ($N_{\rm HB}$, $N_{\rm RR}$, and $N_{\rm BS}$), the \citet{1962AJ.....67..471K} model central concentration $c$, core radius $R_{\rm C}$, half mass radius $R_{\rm HM}$, absolute visual magnitude $M_{\rm V}$, metallicity [Fe/H], HB morphology HBR, distance from the Sun $R_{\rm sun}$, distance from the Galactic center $R_{\rm GC}$, the PCC and AC properties (see Table~\ref{Tab:smpB-mc}).

From Fig.~\ref{Fig:f-par}, we can draw the following conclusions.
(1) Binary fraction $f$ is positively correlated with $M_{\rm V}$, which is consistent with previous conclusion. 
(2) $f_{(R_{\rm C}-R_{\rm HM})}$ and $f_{(0-R_{\rm 1})}$ are not positively correlated to the number of BSs, which is inconsistent with the conclusion drawn by M12 that $f_{(0-R_{\rm C})}$ is positively correlated with $N_{\rm BS}$. 
(3) The number of high binary fraction GCs at HBR=$-1$ is greater than at HBR=1, which maybe is the result that extreme HB stars being formed from binary stars \citep{2002MNRAS.336..449H, 2003MNRAS.341..669H} or the result of the hight binary fraction for dSph galaxies accreted by the MW.
(4) $f$ seems to be positively correlated with $R_{\rm sun}$ and $R_{\rm GC}$, which is caused by observation selection effect. 
(5) $f$ seems to be inversely correlated with $N_{\rm HB}$ and $N_{\rm RR}$, which maybe relates to metallicity. Moreover, $N_{\rm HB}$ and $N_{\rm RR}$ are important indicators of accreted GCs, the relations of $f$$-$$N_{\rm HB}$ and $f$$-$$N_{\rm RR}$ maybe are mislead by the high binary fraction of dSph galaxies accreted by the MW.
(6) $f$ is inversely correlated with $R_{\rm C}$ and $R_{\rm HM}$ (the larger the radii, the smaller the binary fraction), this maybe is due to the overestimation of binary fractions for those dense GCs.

\section{Summary}
\label{sec:summary}
In this paper, we proposed a binary fraction determination method for distant or dense stellar systems by using their integrated properties.
Using the method, the binary fraction variations of 21 GCs (Sample A) first were presented. By comparing the results with those based on the main-sequence fiducial line method, we found that the results in this work can well reproduce previous results. This means that the method proposed in this paper is feasible. It breaks the limitations of traditional methods (see Introduction) and can be used to obtain the binary fractions for dense or distant stellar systems. If we want to obtain more accurate binary fraction and $\Delta$$f$ calibrations in terms of $\Delta$SAFIs, we suggest that the spectroscopic and photometric observations are conduced at an appropriate regional interval based on the properties of stellar systems.

Next, we presented the binary fractions of 13 GCs (Sample B), of which binary fractions have not been studied before. Among these 13 GCs, there is a accreted and post core-collapsed GC, NGC1904, whose binary fraction maybe contains informations about the accretion history of the Milky Way and GC's dynamic evolution. In addition, there are two GCs (NGC6266 and NGC6316) related to binary millisecond pulsars, of which binary fractions can help shape our understanding the formation of binary millisecond pulsars.

By combining the M12 and J15 results, we studied the relationships between binary fraction and various parameters.
The conclusions are listed as follows.
(1) Binary fraction is positively correlated with $M_{\rm V}$, which is consistent with previous studies.
(2) $f_{(R_{\rm C}-R_{\rm HM})}$ and $f_{(0-R_{\rm 1})}$ are inversely correlated with $N_{\rm BS}$, which is inconsistent with the conclusion made by M12 that $f_{(0-R_{\rm C})}$ is positively correlated with $N_{\rm BS}$. 
(3) The fraction of high binary fraction for the GCs with HBR=$-1$ is large, which maybe is the result of the extreme HB binary star model of \citet{2002MNRAS.336..449H, 2003MNRAS.341..669H} or the result of the high binary fraction for dSph galaxies accreted by the MW.
(4) Binary fraction seems to be positively correlated with $R_{\rm sun}$ and $R_{\rm GC}$,  which is caused by an observational selection effect. 
(5) Binary fraction seems to be inversely correlated with $R_{\rm C}$ and $R_{\rm HM}$, which is perhaps because the binary fractions are overestimated for dense star clusters or because the binary fraction is large for dSph galaxies accreted by the MW.

In the subsequent work, we will use Lijiang 2.4-m telescope to perform spectroscopic observations of GCs, improve the $\Delta$$f$ calibration in terms of $\Delta$SAFIs, and determine the binary fractions for some stellar systems in which the individual stars cannot be resolved.

\section*{Acknowledgements}
\label{sec:ack} 
This work was funded by the National Natural Science Foundation of China (No. 11973081), the Science Research Grants from the China Manned Space Project (No. CMS-CSST-2021-A08), 
and the International Centre of Supernovae, Yunnan Key Laboratory (No. 202302AN360001).
We are also grateful to the referee for her/his useful comments. 

\section*{DATA AVAILABILITY}
The data underlying this article is available on reasonable request to the corresponding author or on Zenodo (10.5281/zenodo.11016711).

\bibliographystyle{mnras}
\bibliography{zfh-mn82} 

\bsp    
\label{lastpage}
\end{document}